\documentclass[journal,onecolumn,11pt]{IEEEtran}

\usepackage{graphicx,epsfig,subfigure}
\usepackage{times}
\usepackage{amsmath,amssymb,latexsym}
\usepackage{setspace}
\usepackage{longtable}
\usepackage{multicol}
\usepackage{cite}
\usepackage{color}

\newtheorem{theorem}{Theorem}
\newtheorem{corollary}{Corollary}
\newtheorem{lemma}{Lemma}

\newtheorem{remark}{Remark}

\providecommand{\E}{{\sf E}} 
 
\providecommand{\Tr}{{\sf Tr}} \providecommand{\diag}{{\sf Diag}}

 \providecommand{\Av}{\mathbf{A}}
 \providecommand{\Bv}{\mathbf{B}}
 \providecommand{\Cv}{\mathbf{C}}
 \providecommand{\Dv}{\mathbf{D}}
 \providecommand{\Ev}{\mathbf{E}}
 \providecommand{\Fv}{\mathbf{F}}

 \providecommand{\Gv}{\mathbf{G}}
\providecommand{\hv}{\mathbf{h}} \providecommand{\Hv}{\mathbf{H}}
 \providecommand{\Iv}{\mathbf{I}}

 \providecommand{\Nv}{\mathbf{N}}
 \providecommand{\Mv}{\mathbf{M}}

 \providecommand{\Ov}{\mathbf{O}}
 
 \providecommand{\Qv}{\mathbf{Q}}

 \providecommand{\Sv}{\mathbf{S}}
 
 \providecommand{\Uv}{\mathbf{U}}
 \providecommand{\Vv}{\mathbf{V}}

 \providecommand{\Xv}{\mathbf{X}}
 \providecommand{\Yv}{\mathbf{Y}}
 \providecommand{\Zv}{\mathbf{Z}}

 \providecommand{\Cc}{{\mathcal C}}

 \providecommand{\Rc}{{\mathcal R}}

\providecommand{\Ot}{\widetilde{\mathbf{O}}}

\providecommand{\Nt}{\widetilde{\mathbf{N}}}

\providecommand{\Abv}{\overline{\mathbf{A}}}
\providecommand{\Xbv}{\overline{\mathbf{X}}}

\providecommand{\Ybv}{\overline{\mathbf{Y}}}

\providecommand{\Zbv}{\overline{\mathbf{Z}}}

\providecommand{\Sbv}{\overline{\mathbf{S}}}

\providecommand{\Hbv}{\overline{\mathbf{H}}}

\providecommand{\T}{\intercal}

\begin{document}
\doublespace
\title{Multiple-Input Multiple-Output Gaussian Broadcast Channels with Confidential Messages}

\author{Ruoheng Liu, Tie Liu, H. Vincent Poor, and Shlomo Shamai (Shitz)%
\thanks{This research was supported by the United States National Science Foundation under Grants
CNS-06-25637 and CCF-07-28208, the European Commission in the framework of the
FP7 Network of Excellence in Wireless
Communications NEWCOM++, and the Israel Science Foundation.}%
\thanks{Ruoheng Liu and H. Vincent Poor are with the Department of Electrical Engineering,
Princeton University, Princeton, NJ 08544, USA (e-mail: \{rliu,poor\}@princeton.edu).}%
\thanks{Tie Liu is with the Department of Electrical and Computer Engineering, Texas
A\&M University, College Station, TX 77843, USA (e-mail: tieliu@tamu.edu).}%
\thanks{Shlomo Shamai (Shitz) is with the Department of Electrical Engineering,
Technion-Israel Institute of Technology, Technion City, Haifa 32000,
Israel (e-mail: sshlomo@ee.technion.ac.il).}%
}

\maketitle

\begin{abstract}
This paper considers the problem of secret communication over a two-receiver
multiple-input multiple-output (MIMO) Gaussian broadcast channel. The
transmitter has two independent messages, each of which is intended for one of
the receivers but needs to be kept asymptotically perfectly secret from the
other. It is shown that, surprisingly, under a matrix power constraint both
messages can be simultaneously transmitted at their respective maximal secrecy
rates. To prove this result, the MIMO Gaussian wiretap channel is revisited and
a new characterization of its secrecy capacity is provided via a new coding
scheme that uses artificial noise and random binning.
\end{abstract}


\section{Introduction}\label{sec:INT}
Rapid advances in wireless technology are quickly moving us toward a
pervasively connected world in which a vast array of wireless
devices, from iPhones to biosensors, seamlessly communicate with one
another. The openness of the wireless medium makes wireless
transmission especially susceptible to eavesdropping. Hence,
security and privacy issues have become increasingly critical for
wireless networks. Although wireless technologies are becoming more
and more secure, eavesdroppers are also becoming smarter. Sole
reliance on cryptographic keys in large distributed networks where
terminals can be compromised is no longer sustainable from the
security perspective. Furthermore, in wireless networks, secure
initial key distribution is difficult and, in fact, can be performed
in perfect secrecy only via physical layer techniques. Therefore,
tackling security at the very basic physical layer is of critical
importance.

In this paper, we study the problem of secret communication over the
multiple-input multiple-output (MIMO) Gaussian broadcast channel
with two receivers. The transmitter is equipped with $t$ transmit
antennas, and receiver $k$, $k=1,2$, is equipped with $r_k$ receive
antennas. A discrete-time sample of the channel can be written as
\begin{equation}
\mathbf{Y}_{k}[m] = \mathbf{H}_k\mathbf{X}[m]+\mathbf{Z}_k[m], \quad
k=1,2 \label{eq:Ch1}
\end{equation}
where $\mathbf{H}_k$ is the (real) channel matrix of size $r_{k}
\times t$, and $\{\mathbf{Z}_k[m]\}_m$ is an independent and
identically distributed (i.i.d.) additive vector Gaussian noise
process with zero mean and identity covariance matrix. The channel
input $\{\mathbf{X}[m]\}_m$ is subject to the matrix power
constraint:
\begin{equation}
\frac{1}{n}\sum_{m=1}^{n}\left(\mathbf{X}[m]\mathbf{X}^{\T}[m]\right)
\preceq \mathbf{S} \label{eq:MC}
\end{equation}
where $\mathbf{S}$ is a positive semidefinite matrix, and ``$\preceq$" denotes ``less
than or equal to" in the positive semidefinite ordering between real symmetric matrices.
Note that \eqref{eq:MC} is a rather general power constraint that subsumes many other
important power constraints including the average total and per-antenna power constraints
as special cases.

\begin{figure}[t]
 \centerline{\includegraphics[width=0.6\linewidth,draft=false]{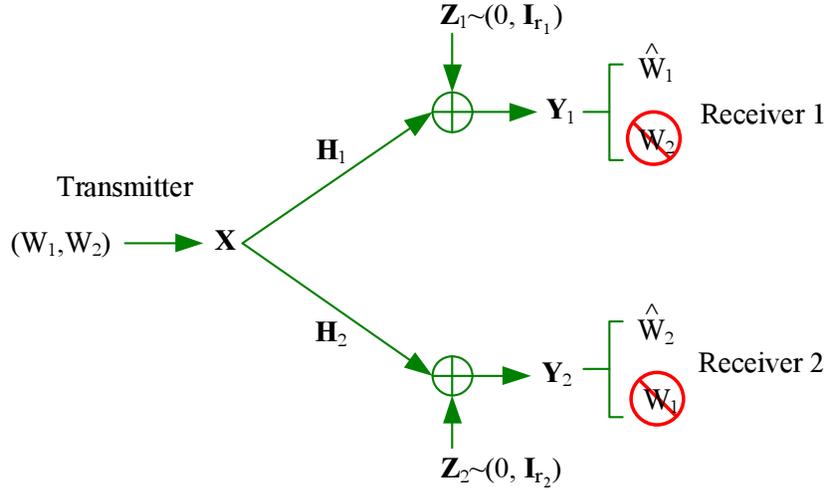}}
 \caption{MIMO Gaussian broadcast channel with confidential messages.}
 \label{fig:gbc}
\end{figure}

Consider the communication scenario in which there are two
independent messages $W_1$ and $W_2$ at the transmitter. Message
$W_1$ is intended for receiver 1 but needs to be kept secret from
receiver 2, and message $W_2$ is intended for receiver 1 but needs
to be kept secret from receiver 2. (See Fig.~\ref{fig:gbc} for an
illustration of this communication scenario.) The confidentiality of
the messages at the unintended receivers is measured using the
normalized information-theoretic quantities
\cite{Wyn-BSTJ75,CK-IT78}:
\begin{equation*}
\frac{1}{n}I(W_1;\mathbf{Y}_{2}^n) \rightarrow 0 \quad \mbox{and}
\quad \frac{1}{n}I(W_2;\mathbf{Y}_{1}^n) \rightarrow 0
\end{equation*}
where $\Yv_k^n:=(\Yv_k[1],\ldots,\Yv_k[n])$, and the limits are taken as the
block length $n \rightarrow \infty$. The goal is to characterize the entire
secrecy rate region $\Cc_s(\Hv_1,\Hv_2,\Sv)=\{(R_1,R_2)\}$ that can be achieved
by any coding scheme. $\Cc_s(\Hv_1,\Hv_2,\Sv)$ is usually known as the
\emph{secrecy capacity region} of the channel.

In recent years, information-theoretic study of secret MIMO communication has
been an active area of research. (See \cite{LPS-book} for a recent survey of
progress in this area.) Most noticeably, the secrecy capacity of the MIMO
Gaussian wiretap channel was characterized in
\cite{Li:CISS:07,KW-ITSubm07,SLU-ITSumb07} for the multiple-input single-output
(MISO) case and \cite{KW-Allerton07,OH-ISIT08,LS-ITSubm07,BLVS-EURASIP08} for
the general MIMO case. The secrecy capacity region of the MIMO Gaussian
broadcast channel with a common and a confidential messages was characterized
in \cite{LLL-ISITA08}. The problem of communicating two confidential messages
over the two-receiver MIMO Gaussian broadcast channel was first considered in
\cite{LP-IT09}, where it was shown that under the average total power
constraint, secret dirty-paper coding (S-DPC) based on double binning
\cite{LMSY-IT08} achieves the secrecy capacity region for the MISO case. For
the general MIMO case, however, characterizing the secrecy capacity region
remained as an open problem.

The main result of this paper is a precise characterization of the secrecy
capacity region of the (general) MIMO Gaussian broadcast channel, summarized in
the following theorem.

\begin{theorem}\label{thm:GMBC}
The secrecy capacity region $\Cc_s(\Hv_1,\Hv_2,\Sv)$ of the MIMO Gaussian
broadcast channel \eqref{eq:Ch1} with confidential messages $W_1$ (intended for
receiver 1 but needing to be kept secret from receiver 2) and $W_2$ (intended
for receiver 2 but needing to be kept secret from receiver 1) under the matrix
power constraint \eqref{eq:MC} is given by the set of nonnegative rate pairs
$(R_1,R_2)$ such that
\begin{align}
R_1 &\le \max_{0\preceq\Bv\preceq\Sv}\left(
\frac{1}{2}\log\left|\Iv_{r_1}+\Hv_1\Bv\Hv_1^{\T}\right|-\frac{1}{2}\log\left|\Iv_{r_2}+\Hv_2
\Bv\Hv_2^{\T}\right|\right)\notag\\
\text{and} \qquad R_2 &\le \max_{0\preceq\Bv\preceq\Sv}\left(
\frac{1}{2}\log\left|\frac{\Iv_{r_2}+\Hv_2\Sv\Hv_2^{\T}}{\Iv_{r_2}+\Hv_2\Bv\Hv_2^{\T}}\right|
-\frac{1}{2}\log\left|\frac{\Iv_{r_1}+\Hv_1\Sv\Hv_1^{\T}}{\Iv_{r_1}+\Hv_1\Bv\Hv_1^{\T}}\right|\right)
\label{eq:SCR}
\end{align}
where $\Iv_{r_k}$ denotes the identity matrix of size $r_k \times
r_k$.
\end{theorem}

\begin{remark}\label{rmk:1}
Note that the rate region \eqref{eq:SCR} is \emph{rectangular}. This
implies that under the matrix power constraint, both confidential
messages $W_1$ and $W_2$ can be \emph{simultaneously} transmitted at
their respective maximal secrecy rates (as if over two separate MIMO
Gaussian wiretap channels). The secrecy capacity of the MIMO
Gaussian wiretap channel under the matrix power constraint was
characterized in \cite{LS-ITSubm07}, by which the rate region
\eqref{eq:SCR} can be rewritten as the set of nonnegative rate pairs
$(R_1,R_2)$ such that
\begin{align}
R_1 &\le \max_{0\preceq\Bv\preceq\Sv}\left(
\frac{1}{2}\log\left|\Iv_{r_1}+\Hv_1\Bv\Hv_1^{\T}\right|-\frac{1}{2}\log\left|\Iv_{r_2}+\Hv_2
\Bv\Hv_2^{\T}\right|\right)\notag\\
\text{and} \qquad R_2 &\le \max_{0\preceq\Bv\preceq\Sv}\left(
\frac{1}{2}\log\left|\Iv_{r_2}+\Hv_2\Bv\Hv_2^{\T}\right|-
\frac{1}{2}\log\left|\Iv_{r_1}+\Hv_1\Bv\Hv_1^{\T}\right|\right).
\label{eq:SCR-ad}
\end{align}
\end{remark}

\begin{remark}\label{rmk:2}
Also note that if $\Bv^\star$ is an optimal solution to the
optimization program:
\begin{align}
\max_{0\preceq\Bv\preceq\Sv}\left(\log\left|\Iv_{r_1}+\Hv_1\Bv\Hv_1^{\T}\right|-\log\left|\Iv_{r_2}+\Hv_2
\Bv\Hv_2^{\T}\right|\right),\label{eq:OPG}
\end{align}
then $\Bv^{\star}$ \emph{simultaneously} maximizes both objective functions on
the right-hand side (RHS) of \eqref{eq:SCR}. On the other hand, the
optimization programs on the RHS of \eqref{eq:SCR-ad} do not, in general, admit
the same optimal solution. As we will see, this makes \eqref{eq:SCR} a better
choice when it comes to proving the achievability part of the theorem.
\end{remark}

It is rather surprising to see that under the matrix power
constraint, both confidential messages $W_1$ and $W_2$ can be
simultaneously transmitted at their respective maximal secrecy rates
over the MIMO Gaussian broadcast channel \eqref{eq:Ch1}. As we will
see, this is due to the fact that there are in fact two different
coding schemes: one uses only random binning, and the other uses
both random binning and \emph{artificial noise}. Both of them can
achieve the secrecy capacity of the MIMO Gaussian wiretap channel.
Through S-DPC (double binning) \cite{LMSY-IT08}, both schemes can be
\emph{simultaneously} implemented in communicating confidential
messages $W_1$ and $W_2$ over the MIMO Gaussian broadcast channel
\eqref{eq:Ch1}.

As a corollary, we have the following characterization of the
secrecy capacity region under the average total power constraint.
The result is a simple consequence of \cite[Lemma~1]{WSS-IT06}.

\begin{corollary}\label{cor:GMBC}
The secrecy capacity region $\Cc_s(\Hv_1,\Hv_2,P)$ of the MIMO Gaussian
broadcast channel \eqref{eq:Ch1} with confidential messages $W_1$ (intended for
receiver 1 but needing to be kept secret from receiver 2) and $W_2$ (intended
for receiver 2 but needing to be kept secret from receiver 1) under the average
total power constraint:
\begin{equation}
\frac{1}{n}\sum_{m=1}^{n}\|\Xv[m]\|^2 \le P \label{eq:ATPC}
\end{equation}
is given by the set of nonnegative rate pairs $(R_1,R_2)$ such that
\begin{align}
R_1 &\le
\frac{1}{2}\log\left|\Iv_{r_1}+\Hv_1\Bv_1\Hv_1^{\T}\right|-\frac{1}{2}\log\left|\Iv_{r_2}+\Hv_2
\Bv_1\Hv_2^{\T}\right|\notag\\
\mbox{and} \quad\quad R_2 &\le
\frac{1}{2}\log\left|\frac{\Iv_{r_2}+\Hv_2(\Bv_1+\Bv_2)\Hv_2^{\T}}{\Iv_{r_2}+\Hv_2\Bv_1\Hv_2^{\T}}\right|-
\frac{1}{2}\log\left|\frac{\Iv_{r_1}+\Hv_1(\Bv_1+\Bv_2)\Hv_1^{\T}}{\Iv_{r_1}+\Hv_1\Bv_1\Hv_1^{\T}}\right|
\label{eq:SCR3}
\end{align}
for some positive semidefinite matrices $\Bv_1$ and $\Bv_2$ such
that $\Tr(\Bv_1+\Bv_2)\le{P}$.
\end{corollary}

\begin{remark}\label{rmk:3}
Unlike Theorem~\ref{thm:GMBC}, under the average total power constraint, the secrecy
capacity region of the MIMO Gaussian broadcast channel is, in general, \emph{not}
rectangular.
\end{remark}

The rest of the paper is devoted to the proof of
Theorem~\ref{thm:GMBC}. As mentioned previously, the rectangular
nature of the rate region \eqref{eq:SCR} suggests that the result is
intimately connected to the secrecy capacity of the MIMO Gaussian
wiretap channel. The secrecy capacity of the MIMO Gaussian wiretap
channel under the matrix power constraint was previously
characterized in \cite{LS-ITSubm07}, where it was shown that
Gaussian random binning \emph{without} prefix coding is optimal. In
Section~\ref{sec:MWTC}, we revisit the MIMO Gaussian wiretap channel
problem and show that Gaussian random binning \emph{with} prefix
coding can also achieve the secrecy capacity, provided that the
prefix channel is appropriately chosen. In Section~\ref{sec:Pf}, we
prove Theorem~\ref{thm:GMBC} using two different characterizations
of the secrecy capacity of the MIMO Gaussian wiretap channel and
S-DPC (double binning) \cite{LMSY-IT08}. Numerical examples are
provided in Section~\ref{sec:Ex} to illustrate the theoretical
results. Finally, in Section~\ref{sec:Con}, we conclude the paper
with some remarks.

\section{MIMO Gaussian Wiretap Channel Revisited}\label{sec:MWTC}
In this section, we revisit the problem of the MIMO Gaussian wiretap
channel under a matrix power constraint. The problem was first
considered in \cite{LS-ITSubm07}, where a precise characterization
of the secrecy capacity was provided. The goal of this section is to
provide an alternative characterization of the secrecy capacity
which will facilitate proving Theorem~\ref{thm:GMBC}. More
specifically, we wish to provide a MIMO wiretap channel bound on the
secrecy rate $R_2$ which will match the RHS of \eqref{eq:SCR}.

\begin{figure}[t]
 \centerline{\includegraphics[width=0.6\linewidth,draft=false]{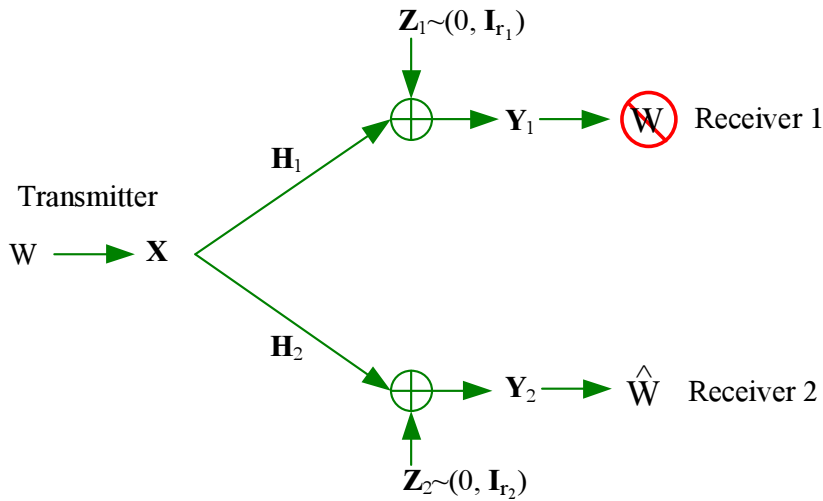}}
 \caption{MIMO Gaussian wiretap channel.}
 \label{fig:WT}
\end{figure}

For that purpose, consider again the MIMO Gaussian broadcast channel
\eqref{eq:Ch1} but this time with only one confidential message $W$
at the transmitter. Message $W$ is intended for receiver 2 (the
legitimate receiver) but needs to be kept secret from receiver 1
(the eavesdropper). The confidentiality of $W$ at receiver 1 is
measured using the normalized information-theoretic quantity
\cite{Wyn-BSTJ75,CK-IT78}:
\begin{equation*}
\frac{1}{n}I(W;\mathbf{Y}_{1}^n) \rightarrow 0.
\end{equation*}
The channel input $\{\Xv[m]\}_m$ is subject to the matrix power
constraint \eqref{eq:MC}. The goal is to characterize the secrecy
capacity $C_s(\Hv_2,\Hv_1,\Sv)$\footnote{In our notation, the first
argument in $C_s(\cdot)$ represents the channel matrix for the
legitimate receiver, and the second argument represents the channel
matrix for the eavesdropper.}, which is the maximum achievable
secrecy rate for message $W$. This communication scenario, as
illustrated in Fig.~\ref{fig:WT}, is widely known as the MIMO
Gaussian wiretap channel
\cite{Li:CISS:07,SLU-ITSumb07,KW-ITSubm07,KW-Allerton07,OH-ISIT08,LS-ITSubm07}.

In their seminal work \cite{CK-IT78}, Csisz\'{a}r and K\"{o}rner
provided a single-letter characterization of the secrecy capacity:
\begin{align}
C_s(\Hv_2,\Hv_1,\Sv)=\max_{(U,\Xv)}\left[I(U;\Yv_2)-I(U;\Yv_1)\right]
\label{eq:SC-CK}
\end{align}
where $U$ is an auxiliary variable, and the maximization is over all jointly distributed
$(U,\Xv)$ such that $U \rightarrow \Xv \rightarrow (\Yv_1,\Yv_2)$ forms a Markov chain
and $\E[\Xv\Xv^\T]\preceq\Sv$. Here, $I(U,\Yv_k)$ denotes the mutual information between
$U$ and $\Yv_k$. As shown in \cite{CK-IT78}, the secrecy rate on the RHS of
\eqref{eq:SC-CK} can be achieved by a coding scheme that combines random binning and
prefix coding \cite{CK-IT78}. More specifically, the auxiliary variable $U$ represents a
precoding signal, and the conditional distribution of $\Xv$ given $U$ represents the
prefix channel. In \cite{LS-ITSubm07}, Liu and Shamai further studied the optimization
problem on the RHS of \eqref{eq:SC-CK} and showed that a Gaussian $U=\Xv$ is an optimal
solution. Hence, a matrix characterization of the secrecy capacity is given by
\cite{LS-ITSubm07}
\begin{align}
C_s(\Hv_2,\Hv_1,\Sv)=\max_{0\preceq\Bv\preceq\Sv}\left(
\frac{1}{2}\log\left|\Iv_{r_2}+\Hv_2\Bv\Hv_2^{\T}\right|-
\frac{1}{2}\log\left|\Iv_{r_1}+\Hv_1\Bv\Hv_1^{\T}\right|\right).
\label{eq:SC-LS}
\end{align}
We may conclude that Gaussian random binning \emph{without} prefix
coding is an optimal coding strategy for the MIMO Gaussian wiretap
channel.

Next, we show that a different coding scheme that combines Gaussian
random binning \emph{and} prefix coding can also achieve the secrecy
capacity of the MIMO Gaussian wiretap channel. This leads to a new
characterization of the secrecy capacity, summarized in the
following theorem.

\begin{theorem}\label{thm:MWTC}
The secrecy capacity $C_s(\Hv_2,\Hv_1,\Sv)$ of the MIMO Gaussian
broadcast channel \eqref{eq:Ch1} with a confidential message $W$
(intended for receiver 2 but needing to be kept secret from receiver
1) under the matrix power constraint \eqref{eq:MC} is given by:
\begin{align}
C_s(\Hv_2,\Hv_1,\Sv) = \max_{0\preceq\Bv\preceq\Sv}\left(
\frac{1}{2}\log\left|\frac{\Iv_{r_2}+\Hv_2\Sv\Hv_2^{\T}}{\Iv_{r_2}+\Hv_2\Bv\Hv_2^{\T}}\right|
-\frac{1}{2}\log\left|\frac{\Iv_{r_1}+\Hv_1\Sv\Hv_1^{\T}}{\Iv_{r_1}+\Hv_1\Bv\Hv_1^{\T}}\right|\right).
\label{eq:SC-LLPS}
\end{align}
\end{theorem}

\begin{remark}\label{rmk:4}
The achievability of the secrecy rate on the RHS of
\eqref{eq:SC-LLPS} can be obtained from the Csisz\'{a}r-K\"{o}rner
expression \eqref{eq:SC-CK} by choosing $\Xv=U+V$, where $U$ and $V$
are two independent Gaussian vectors with zero means and covariance
matrices $\Sv-\Bv$ and $\Bv$, respectively. This choice of $(U,\Xv)$
differs from that for \eqref{eq:SC-LS} in two important ways:
\begin{enumerate}
\item In \eqref{eq:SC-LLPS}, the input vector $\Xv$ always has a full
covariance matrix $\Sv$. For \eqref{eq:SC-LS}, the covariance matrix of $\Xv$
needs to be chosen to solve an optimization program; the full covariance matrix
$\Sv$ is \emph{not} always an optimal solution.
\item In \eqref{eq:SC-LLPS}, the conditional distribution of $\Xv$ given
$U$ may form a \emph{nontrivial} prefix channel. For \eqref{eq:SC-LS},
$U\equiv\Xv$ so prefix coding is never applied.
\end{enumerate}
\end{remark}

\begin{remark}\label{rmk:5}
Note that the prefix channel in \eqref{eq:SC-LLPS} is an additive vector
Gaussian noise channel, so the auxiliary variable $V$ represents an
\emph{artificial} noise \cite{GN-WC08} sent (on purpose) by the transmitter to
confuse the eavesdropper. Since the artificial noise has no structure to it, it
will add to the noise floor at both legitimate receiver and the eavesdropper.
\end{remark}

The converse part of the theorem can be proved using a
\emph{channel-enhancement} argument, similar to that in
\cite{LS-ITSubm07}. The details of the proof are provided in
Appendix~\ref{app:pf-MWTC}.

\section{MIMO Gaussian Broadcast Channel with Confidential Messages}\label{sec:Pf}
In this section, we prove Theorem~\ref{thm:GMBC}. To prove the converse part of
the theorem, we will consider a single-message, wiretap channel bound on the
secrecy rates $R_1$ and $R_2$. More specifically, note that both messages $W_1$
and $W_2$ can be transmitted at the maximum secrecy rate when the other message
is absent from the transmission. Therefore, to bound from above the secrecy
rate $R_1$, we assume that only $W_1$ needs to be communicated over the
channel. This is precisely a MIMO Gaussian wiretap channel problem with
receiver 1 as legitimate receiver and receiver 2 as eavesdropper. Reversing the
roles of receiver 1 and 2, we have from (\ref{eq:SC-LS}) that
\begin{align}
R_1 &\le C_s(\Hv_1,\Hv_2,\Sv)\notag\\
&= \max_{0\preceq\Bv\preceq\Sv}\left(
\frac{1}{2}\log\left|\Iv_{r_1}+\Hv_1\Bv\Hv_1^{\T}\right|-
\frac{1}{2}\log\left|\Iv_{r_2}+\Hv_2\Bv\Hv_2^{\T}\right|\right).\label{eq:R1a}
\end{align}
Similarly, to bound from above the secrecy rate $R_2$, let us assume that only $W_2$
needs to be communicated over the channel. This is, again, a MIMO Gaussian wiretap
channel problem with receiver 2 playing the role of legitimate receiver and receiver 1
playing the role of eavesdropper. By Theorem~\ref{thm:MWTC},
\begin{align}
R_2 &\le C_s(\Hv_2,\Hv_1,\Sv)\notag\\
&= \max_{0\preceq\Bv\preceq\Sv}\left(
\frac{1}{2}\log\left|\frac{\Iv_{r_2}+\Hv_2\Sv\Hv_2^{\T}}{\Iv_{r_2}+\Hv_2\Bv\Hv_2^{\T}}\right|
-\frac{1}{2}\log\left|\frac{\Iv_{r_1}+\Hv_1\Sv\Hv_1^{\T}}{\Iv_{r_1}+\Hv_1\Bv\Hv_1^{\T}}\right|\right).
\label{eq:R2a}
\end{align}
Putting together \eqref{eq:R1a} and \eqref{eq:R2a}, we have proved
the converse part of the theorem.

Next, we show that every rate pair $(R_1,R_2)$ within the secrecy
rate region \eqref{eq:SCR} is achievable. Note that \eqref{eq:SCR}
is rectangular, so we only need to show that the corner point
$(R_1,R_2)$ given by
\begin{align}
R_1 &=\max_{0\preceq\Bv\preceq\Sv}
\left(\frac{1}{2}\log\left|\Iv_{r_1}+\Hv_1\Bv\Hv_1^{\T}\right|-
\frac{1}{2}\log\left|\Iv_{r_2}+\Hv_2
\Bv\Hv_2^{\T}\right|\right)\notag\\
\text{and}\qquad R_2
&=\max_{0\preceq\Bv\preceq\Sv}\left(\frac{1}{2}\log\left|\frac{\Iv_{r_2}+\Hv_2\Sv\Hv_2^{\T}}{\Iv_{r_2}+\Hv_2\Bv\Hv_2^{\T}}\right|
-\frac{1}{2}\log\left|\frac{\Iv_{r_1}+\Hv_1\Sv\Hv_1^{\T}}{\Iv_{r_1}+\Hv_1\Bv\Hv_1^{\T}}\right|\right)
\label{eq:CP}
\end{align}
is achievable.

Recall from \cite{LMSY-IT08} that for any jointly distributed
$(V_1,V_2,\Xv)$ such that
$(V_1,V_2)\rightarrow\Xv\rightarrow(\Yv_1,\Yv_2)$ forms a Markov
chain and $\E[\Xv\Xv^\T]\preceq\Sv$, the secrecy rate pair
$(R_1,R_2)$ given by
\begin{align}
R_1 &=
I(V_1;\Yv_1)-I(V_1;V_2,\Yv_2)\notag\\
\text{and}\qquad R_2 &=I(V_2;\Yv_2)-I(V_2;V_1,\Yv_1) \label{eq:ASRR-LP}
\end{align}
is achievable for the MIMO Gaussian broadcast channel \eqref{eq:Ch1} under the
matrix power constraint \eqref{eq:MC}. In \cite{LMSY-IT08}, the achievability
of the rate pair \eqref{eq:ASRR-LP} was proved using a \emph{double-binning}
scheme. Specifically, the auxiliary variables $V_1$ and $V_2$ represent the
precoding signals for the confidential messages $W_1$ and $W_2$, respectively.

Now let $\Bv$ be a positive semidefinite matrix such that
$\Bv\preceq\Sv$, and let
\begin{align}
V_1 &=\Uv_1+\Fv \Uv_2\notag\\
V_2 &=\Uv_2\notag\\
\mbox{and} \quad \quad \Xv &=\Uv_1+\Uv_2 \label{eq:RVs}
\end{align}
where $\Uv_1$ and $\Uv_2$ are two independent Gaussian vectors with zero means and
covariance matrices $\Bv$ and $\Sv-\Bv$, respectively, and
\begin{align}
\Fv &:=\Bv\Hv_1^{\T}(\Iv_{r_1}+\Hv_1\Bv\Hv_1^{\T})^{-1}\Hv_1. \label{eq:F-def}
\end{align}
By \eqref{eq:RVs}, $$\Yv_k=\Hv_k(\Uv_1+\Uv_2)+\Zv_k$$ for $k=1,2$.
Note that the matrix $\Fv$ defined in \eqref{eq:F-def} is precisely
the \emph{precoding} matrix for suppressing $\Uv_2$ from $\Yv_1$
\cite[Theorem~1]{Yu:IT:04}. Hence,
\begin{align}
I(V_1;\Yv_1)-I(V_1;V_2)
&=I(V_1;\Yv_1)-I(V_1;\Uv_2)\notag\\
&=\frac{1}{2}\log\left|\Iv_{r_1}+\Hv_1\Bv\Hv_1^{\T}\right|.
\label{eq:DPC1}
\end{align}
Moreover,
\begin{align}
I(V_1;\Yv_2|V_2) &=I(\Uv_1+\Fv\Uv_2;\Hv_2(\Uv_1+\Uv_2)+\Zv_2|\Uv_2)\notag\\
&=I(\Uv_1;\Hv_2\Uv_1+\Zv_2|\Uv_2)\notag\\
&=I(\Uv_1;\Hv_2\Uv_1+\Zv_2)\notag\\
&=\frac{1}{2}\log\left|\Iv_{r_2}+\Hv_2\Bv\Hv_2^{\T}\right|
\label{eq:DPC2}
\end{align}
where the third equality follows from the fact that $\Uv_1$ and
$\Uv_2$ are independent. Putting together (\ref{eq:DPC1}) and
(\ref{eq:DPC2}), we have
\begin{align}
I(V_1;\Yv_1)-I(V_1;V_2,\Yv_2) &=
[I(V_1;\Yv_1)-I(V_1;V_2)]-I(V_1;\Yv_2|V_2)\notag\\
&=\frac{1}{2}\log\left|\Iv_{r_1}+\Hv_1\Bv\Hv_1^{\T}\right|-
\frac{1}{2}\log\left|\Iv_{r_2}+\Hv_2\Bv\Hv_2^{\T}\right|.\label{eq:R1}
\end{align}
Similarly,
\begin{align}
I(V_1,V_2;\Yv_1) &=
I(\Uv_1+\Fv\Uv_2,\Uv_2;\Hv_1(\Uv_1+\Uv_2)+\Zv_2)\notag\\
&= I(\Uv_1,\Uv_2;\Hv_1(\Uv_1+\Uv_2)+\Zv_2)\notag\\
&= \frac{1}{2}\log\left|\Iv_{r_1}+\Hv_1\Sv\Hv_1^{\T}\right|.
\label{eq:DPC3}
\end{align}
Thus,
\begin{align}
I(V_2;V_1,\Yv_1) &=I(V_2;\Yv_1|V_1)+I(V_2;V_1)\notag\\
&= I(V_1,V_2;\Yv_1)-[I(V_1;\Yv_1)-I(V_1;V_2)]\notag\\
&= \frac{1}{2}\log\left|\frac{\Iv_{r_1}+\Hv_1\Sv\Hv_1^{\T}}
{\Iv_{r_1}+\Hv_1\Bv\Hv_1^{\T}}\right| \label{eq:DPC4}
\end{align}
where the last equality follows from \eqref{eq:DPC1} and \eqref{eq:DPC3}. Moreover,
\begin{align}
I(V_2;\Yv_2) &= I(\Uv_2;\Hv_2(\Uv_1+\Uv_2)+\Zv_2)\notag\\
&= \frac{1}{2}\log\left|\frac{\Iv_{r_2}+\Hv_2\Sv\Hv_2^{\T}}
{\Iv_{r_2}+\Hv_2\Bv\Hv_2^{\T}}\right|. \label{eq:DPC5}
\end{align}
Putting together (\ref{eq:DPC4}) and (\ref{eq:DPC5}), we have
\begin{align}
I(V_2;\Yv_2)-I(V_2;V_1,\Yv_1) &=
\frac{1}{2}\log\left|\frac{\Iv_{r_2}+\Hv_2\Sv\Hv_2^{\T}}
{\Iv_{r_2}+\Hv_2\Bv\Hv_2^{\T}}\right|-
\frac{1}{2}\log\left|\frac{\Iv_{r_1}+\Hv_1\Sv\Hv_1^{\T}}
{\Iv_{r_1}+\Hv_1\Bv\Hv_1^{\T}}\right|.\label{eq:R2}
\end{align}

Finally, let $\Bv$ be an optimal solution to the optimization
program \eqref{eq:OPG}. As mentioned previously in
Remark~\ref{rmk:2}, such a choice will \emph{simultaneously}
maximize the RHS of \eqref{eq:R1} and \eqref{eq:R2}. Thus, the
corner point \eqref{eq:CP} is indeed achievable. This completes the
proof of the theorem.

\begin{remark}\label{rmk:6}
Note that in standard dirty-paper coding (DPC), the precoding matrix
$\Fv$ is chosen to cancel the known interference. In our scheme,
such a choice plays two important roles. First, it helps to cancel
the precoding signal representing message $W_2$, so message $W_1$
sees an interference-free legitimate receiver channel. Second, it
helps to boost the security for message $W_2$ by causing
interference to its eavesdropper. For this reason, we call our
scheme S-DPC, to differentiate from the standard DPC.
\end{remark}

\begin{remark}\label{rmk:7}
In S-DPC, both the legitimate receiver and the eavesdropper for
message $W_1$ are interference free. On the other hand, for message
$W_2$, both the legitimate receiver and the eavesdropper are subject
to interference from the precoding signal representing message
$W_1$. As we have seen in Section~\ref{sec:MWTC}, the secrecy
capacity of the MIMO Gaussian wiretap channel can be achieved with
or without interference in place. Therefore, both secrecy capacity
achieving schemes can be simultaneously implemented via S-DPC to
simultaneously communicate both confidential messages at their
respective maximal secrecy rates.
\end{remark}

\section{Numerical Examples}\label{sec:Ex}
In this section, we provide numerical examples to illustrate the
secrecy capacity region of the MIMO Gaussian wiretap channel with
confidential messages. As shown in \eqref{eq:SCR} and
\eqref{eq:SCR3}, under both matrix and average total power
constraints, the secrecy capacity regions $\Cc_s(\Hv_1,\Hv_2,\Sv)$
and $\Cc_s(\Hv_1,\Hv_2,P)$ are expressed in terms of matrix
optimization programs (though implicit in \eqref{eq:SCR3}). In
general, these optimization programs are not convex, and hence,
finding the boundary of the secrecy capacity regions is nontrivial.

In \cite{LP-IT09}, a precise characterization of the secrecy
capacity region $\Cc_s(\Hv_1,\Hv_2,P)$ was obtained for the
\emph{MISO} Gaussian broadcast channel using the generalized
eigenvalue decomposition \cite[Ch.~6.3]{Gilbert}. For the
\emph{aligned} MIMO Gaussian wiretap channel, \cite{BLVS-EURASIP08}
provided an explicit, closed-form expression for the secrecy
capacity. In the following, we generalize the results of
\cite{BLVS-EURASIP08} and \cite{LP-IT09} to the general MIMO
Gaussian broadcast channel under the matrix power constraint.

Let $\phi_j$, $j=1,\ldots,t$, be the generalized eigenvalues of the
pencil
\begin{align}
\left(\Iv_t+\Sv^{\frac{1}{2}}\Hv_1^{\T}\Hv_1\Sv^{\frac{1}{2}},\;
\Iv_t+\Sv^{\frac{1}{2}}\Hv_2^{\T}\Hv_2\Sv^{\frac{1}{2}}\right).
\label{eq:Pen}
\end{align}
Since both $\Iv_t+\Sv^{\frac{1}{2}}\Hv_1^{\T}\Hv_1
\Sv^{\frac{1}{2}}$ and
$\Iv_t+\Sv^{\frac{1}{2}}\Hv_2^{\T}\Hv_2\Sv^{\frac{1}{2}}$ are
strictly positive definite, we have $\phi_{j}>0$ for $j=1,\dots,t$.
Without loss of generality, we may assume that these generalized
eigenvalues are ordered as
$$\phi_{1}\ge\dots\ge \phi_{\rho}> 1 \ge \phi_{\rho+1}\ge\dots\ge \phi_{t}>0,$$
i.e., a total of $\rho$ of them are assumed to be greater than $1$.
We have the following characterization of the secrecy capacity of
the MIMO Gaussian wiretap channel under the matrix power constraint,
which is a natural extension of \cite{BLVS-EURASIP08}.

\begin{theorem}\label{thm:MWTC2}
The secrecy capacity $C_s(\Hv_1,\Hv_2,\Sv)$ of the MIMO Gaussian
broadcast channel \eqref{eq:Ch1} with confidential message $W$
(intended for receiver 1 but needing to be kept secret from receiver
2) under the matrix power constraint \eqref{eq:MC} is given by
\begin{align}
C_s(\Hv_1,\Hv_2,\Sv) = \frac{1}{2}\sum_{j=1}^{\rho}\log\phi_j
\label{eq:MWTC2}
\end{align}
where $\phi_j$, $j=1,\ldots,\rho$, are the generalized eigenvalues
of the pencil \eqref{eq:Pen} that are greater than 1.
\end{theorem}

\begin{remark}
Note that $\Iv_t+\Sv^{\frac{1}{2}}\Hv_2^{\T}\Hv_2\Sv^{\frac{1}{2}}$
is invertible, so computing the generalized eigenvalues of the
pencil \eqref{eq:Pen} can be reduced to the problem of finding
standard eigenvalues of a related semidefinite matrix
\cite[Ch.~6.3]{Gilbert}. Hence, the secrecy capacity expression
\eqref{eq:MWTC2} is computable.
\end{remark}

A proof of the theorem following the approach of
\cite{BLVS-EURASIP08} is provided in Appendix~\ref{app:pf-MWTC2}. As
a corollary, we have the following characterization of the secrecy
capacity region of the MIMO Gaussian broadcast channel with
confidential messages under the matrix power constraint.

\begin{corollary}\label{cor:GMBC2}
The secrecy capacity region $\Cc_s(\Hv_1,\Hv_2,\Sv)$ of the MIMO Gaussian
broadcast channel \eqref{eq:Ch1} with confidential messages $W_1$ (intended for
receiver 1 but needing to be kept secret from receiver 2) and $W_2$ (intended
for receiver 2 but needing to be kept secret from receiver 1) under the matrix
constraint \eqref{eq:MC} is given by the set of nonnegative rate pairs
$(R_1,R_2)$ such that
\begin{align}
R_1 &\le \frac{1}{2}\sum_{j=1}^{\rho}\log\phi_j\notag\\
\text{and} \qquad R_2 &\le
\frac{1}{2}\sum_{j=\rho+1}^{t}\log\frac{1}{\phi_j}
\label{eq:SCR2}
\end{align}
where $\phi_j$, $j=1,\ldots,\rho$, are the generalized eigenvalues
of the pencil \eqref{eq:Pen} that are greater than 1, and $\phi_j$,
$j=\rho+1,\ldots,t$, are the generalized eigenvalues of the pencil
\eqref{eq:Pen} that are less than or equal to 1.
\end{corollary}

\begin{proof}
By Theorem~\ref{thm:GMBC}, we only need to show that the secrecy
capacity
\begin{align*}
C_s(\Hv_2,\Hv_1,\Sv) =
\frac{1}{2}\sum_{j=\rho+1}^{t}\log\frac{1}{\phi_j}.
\end{align*}
Consider the pencil
\begin{align}
\left(\Iv_t+\Sv^{\frac{1}{2}}\Hv_2^{\T}\Hv_2\Sv^{\frac{1}{2}},\;
\Iv_t+\Sv^{\frac{1}{2}}\Hv_1^{\T}\Hv_1\Sv^{\frac{1}{2}}\right).
\label{eq:Pen2}
\end{align}
Note that the pencils \eqref{eq:Pen} and \eqref{eq:Pen2} are generated by the same pair
of semidefinite matrices but with different order. Therefore, the generalized eigenvalues
of the pencil \eqref{eq:Pen2} are given by
$$0<\frac{1}{\phi_{1}}\le\dots \le \frac{1}{\phi_{\rho}}<1
\le \frac{1}{\phi_{\rho+1}}\le\dots\le\frac{1}{\phi_{t}}.$$ Applying
Theorem~\ref{thm:MWTC2} for $C_s(\Hv_2,\Hv_1,\Sv)$ completes the
proof of the corollary.
\end{proof}

Under the average total power constraint, we have not been able to
find a computable secrecy capacity expression for the general MIMO
case. We can, however, write \cite[Lemma~1]{WSS-IT06}
$$\Cc_s(\Hv_1,\Hv_2,P)=\bigcup_{\Sv\succeq0,\;\Tr(\Sv)\le
P}\Cc_s(\Hv_1,\Hv_2,\Sv).$$ For any given semidefinite $\Sv$,
$\Cc_s(\Hv_1,\Hv_2,\Sv)$ can be computed as given by
\eqref{eq:SCR2}. Then, the secrecy capacity region
$\Cc_s(\Hv_1,\Hv_2,P)$ can be found through an exhaustive search
over the set $\{\Sv:\;\Sv\succeq0\;\mbox{and}\;\Tr(\Sv)\le P\}$.

\begin{figure}[t]
\begin{minipage}[b]{0.5\linewidth}
\centerline{\includegraphics[width=\linewidth,draft=false]{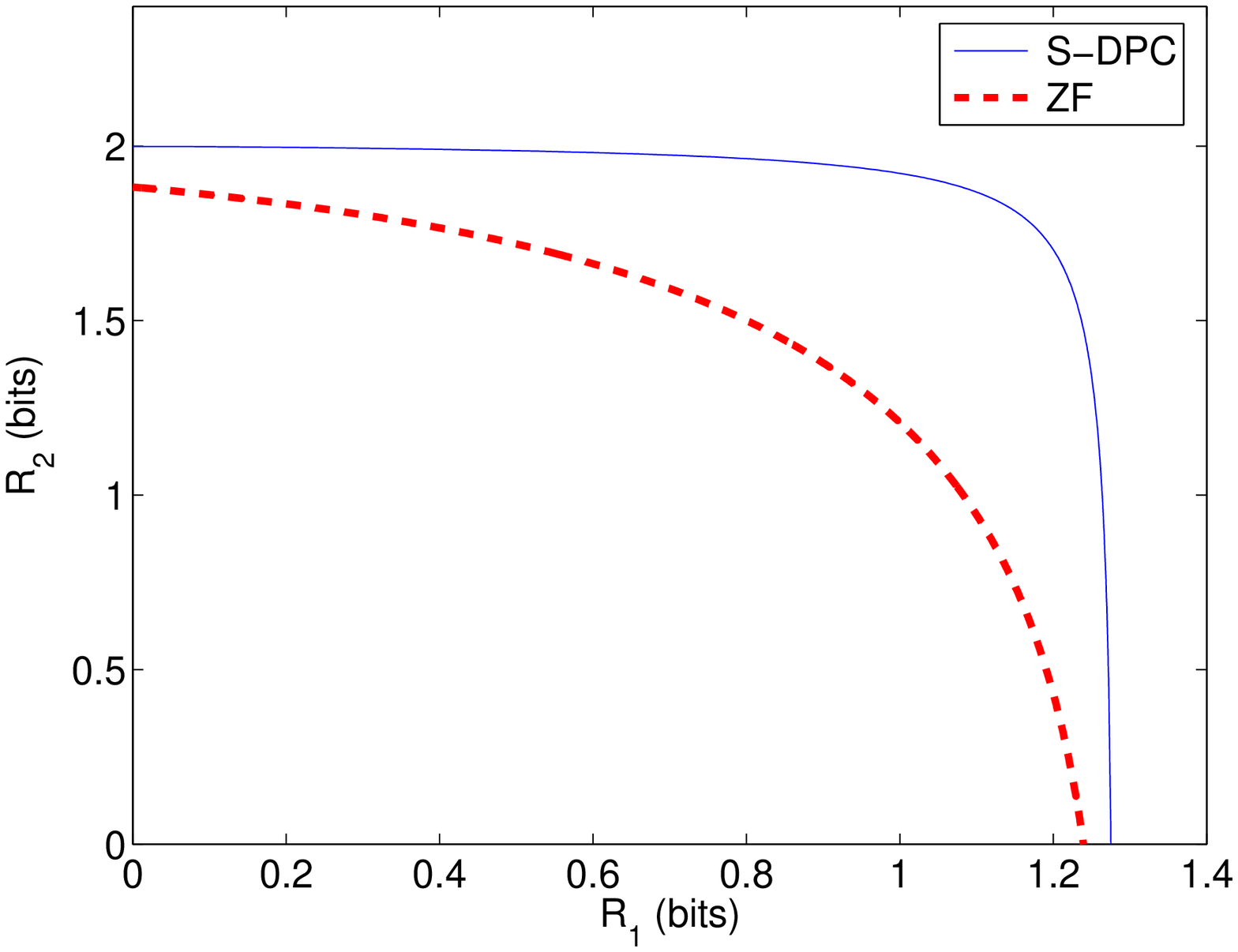}}
\centerline{\mbox{\footnotesize (a) $r_1=r_2=1$}} 
\end{minipage}\hfill
\begin{minipage}[b]{0.5\linewidth}
\centerline{\includegraphics[width=\linewidth,draft=false]{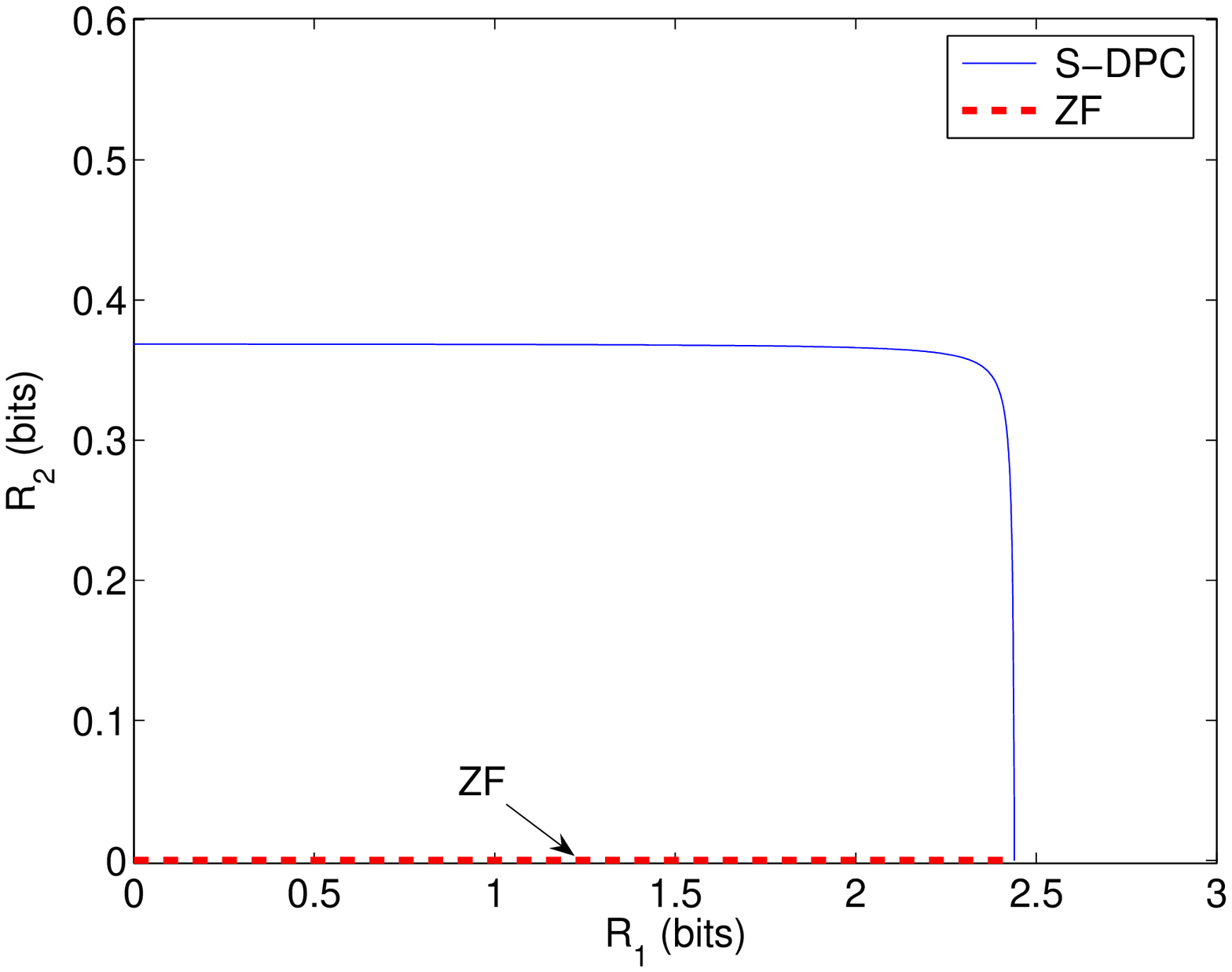}}
\centerline{\mbox{\footnotesize (b) $r_1=2$, $r_2=1$}} 
\end{minipage}
\begin{minipage}[b]{0.5\linewidth}
\centerline{\includegraphics[width=\linewidth,draft=false]{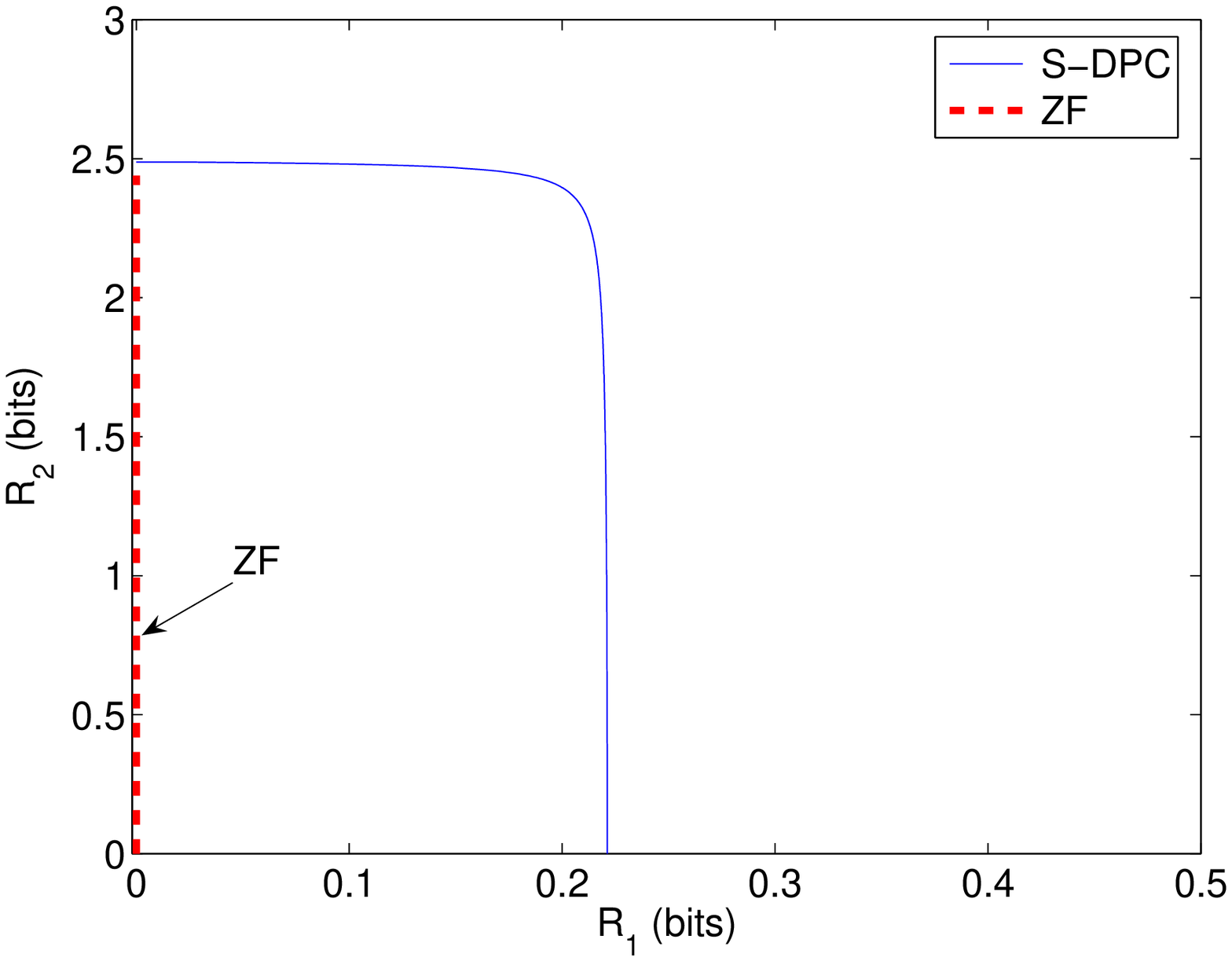}}
\centerline{\mbox{\footnotesize (c) $r_1=1$, $r_2=2$}}
\end{minipage}\hfill
\begin{minipage}[b]{0.5\linewidth}
\centerline{\includegraphics[width=\linewidth,draft=false]{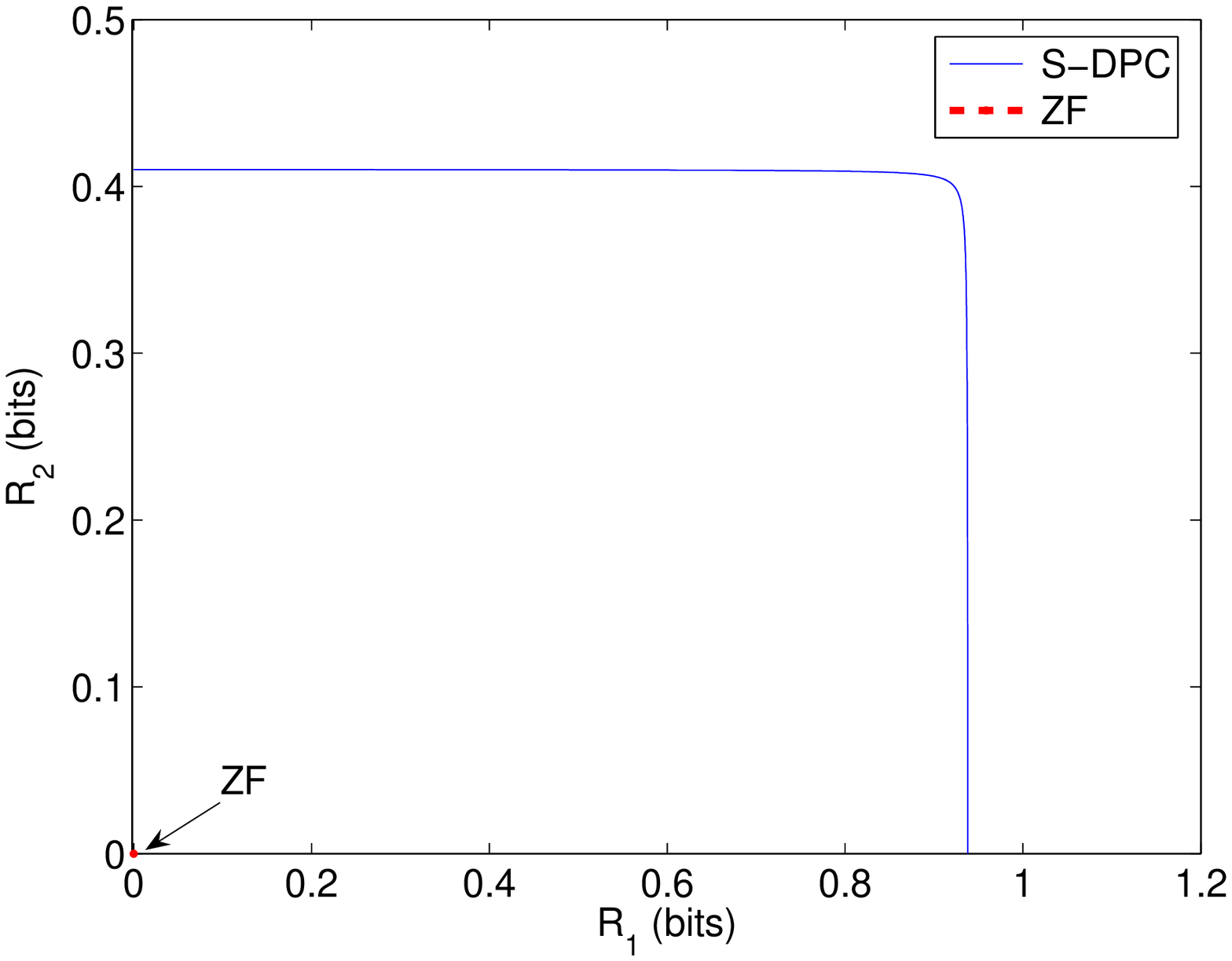}}
\centerline{\mbox{\footnotesize (d) $r_1=r_2=2$}}
\end{minipage}
\caption {Secrecy rate regions of the MIMO Gaussian broadcast channel under the
average total power constraint.} \label{fig:sim}
\end{figure}

Let $\hv_{11}=(0.3\; 2.5)$, $\hv_{12}=(2.2\; 1.8)$, $\hv_{21}=(1.3\; 1.2)$,
$\hv_{22}=(1.5\; 3.9)$ and $P=12$, and let
\begin{align*}
\Hv_k &= \left(
          \begin{array}{c}
            \hv_{k1} \\
            \hv_{k2} \\
          \end{array}
        \right), \quad k=1,2.
\end{align*}
The secrecy capacity regions $\Cc_s(\hv_{11},\hv_{22},P)$,
$\Cc_s(\Hv_1,\hv_{22},P)$, $\Cc_s(\hv_{11},\Hv_2,P)$ and $\Cc_s(\Hv_1,\Hv_2,P)$
are illustrated in Fig.~\ref{fig:sim}. For comparison, we have also plotted the
secrecy rate regions achieved by the simple zero-forcing (ZF) strategy. In ZF,
each of the confidential messages is encoded using a vector Gaussian signal. To
guarantee confidentiality, the covariance matrices of the transmit signals are
chosen in the \emph{null} space of the channel matrix at the unintended
receiver. Hence, the achievable secrecy rate region is given by
\begin{align}
\Rc_S^{\rm ZF}(\Hv_{1},\Hv_{2},P)=\bigcup_{\substack{\Bv_1\succeq
0,\;\Bv_2\succeq 0,\; \Tr(\Bv_1+\Bv_2)\le P
\\\Hv_2\Bv_1=0,\;
 \Hv_1\Bv_2=0}} \left\{(R_1,R_2) \left|\;
\begin{array}{l}
R_1\le \frac{1}{2}\log|\Iv_{r_1}+\Hv_1\Bv_1\Hv_1^{\T}|\\
R_2\le \frac{1}{2}\log|\Iv_{r_2}+\Hv_2\Bv_2\Hv_2^{\T}|
\end{array}
\right. \right\}. \label{eq:ZF}
\end{align}
Note that unlike the secrecy capacity region expression
\eqref{eq:SCR3}, computing the rate region \eqref{eq:ZF} only
involves solving convex optimization programs. As shown in
Fig.~\ref{fig:sim}, in all four scenarios, ZF is strictly suboptimal
as compared with S-DPC. In particular, if the channel matrix of the
unintended receiver has full row rank, ZF cannot achieve any
positive secrecy rate for the corresponding confidential message. On
the other hand, S-DPC can always achieve positive secrecy rates for
both confidential messages unless the MIMO Gaussian broadcast
channel is degraded.

\begin{figure}[t]
\centerline{\includegraphics[width=0.5\linewidth,draft=false]{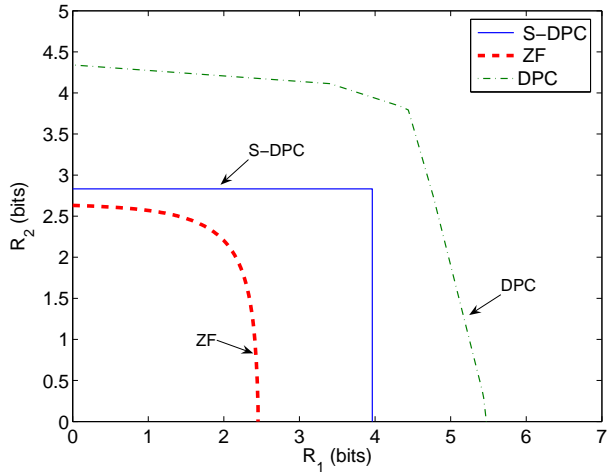}}
\caption{Rate regions of the MIMO Gaussian broadcast channel under the power
matrix constraint.} \label{fig:sim2}
\end{figure}

Finally, let
\begin{align*}
\Hv_1 &=\left(\begin{matrix} 1.8&  -2.0 & 2.0 \\ 1.0 &  -6.0 & 3.0
\end{matrix}\right)\\
\Hv_2 &=\left(\begin{matrix} 2.3 & 2.0 & -3 \\
2.0 & 1.2& -1.5
\end{matrix}\right)
\end{align*}
and
\begin{align*}
\Sv=\left(\begin{matrix} 5.0 & -0.7 & -2.0 \\ -0.7 & 3.8 & -2.5 \\
-2.0 & -2.5 & 5.0 \end{matrix}\right).
\end{align*}
Fig.~\ref{fig:sim2} illustrates the secrecy capacity region
$\mathcal{C}_s(\Hv_1,\Hv_2,\Sv)$ of the MIMO Gaussian broadcast
channel \eqref{eq:Ch1} under the matrix power constraint
\eqref{eq:MC}. Here, the secrecy capacity region
$\mathcal{C}_s(\Hv_1,\Hv_2,\Sv)$ is plotted based on the computable
expression \eqref{eq:SCR2}. Also in the figure are the secrecy rate
region $\mathcal{R}_s^{\rm ZF}(\Hv_1,\Hv_2,\Sv)$ achieved by ZF
strategy and the \emph{nonsecrecy} capacity region $\mathcal{R}^{\rm
DPC}(\Hv_1,\Hv_2,\Sv)$ achieved by standard DPC \cite{WSS-IT06}. As
expected, we have $\mathcal{R}_s^{\rm ZF}(\Hv_1,\Hv_2,\Sv)\subset
\mathcal{C}_s(\Hv_1,\Hv_2,\Sv)\subset \mathcal{R}^{\rm
DPC}(\Hv_1,\Hv_2,\Sv)$.

\section{Concluding Remarks} \label{sec:Con}
In this paper, we have considered the problem of communicating two
confidential messages over the two-receiver MIMO Gaussian broadcast
channel. Each of the confidential messages is intended for one of
the receivers but needs to be kept asymptotically perfectly secret
from the other. Precise characterizations of the secrecy capacity
region have been provided under both matrix and average total power
constraints. Surprisingly, under the matrix power constraint, both
confidential messages can be transmitted simultaneously at their
respective maximal secrecy rates.

To prove this result, we have revisited the problem of the MIMO
Gaussian wiretap channel and proposed a new coding scheme that
achieves the secrecy capacity of the channel. Unlike the previous
scheme considered in
\cite{Li:CISS:07,SLU-ITSumb07,KW-ITSubm07,KW-Allerton07,OH-ISIT08,LS-ITSubm07}
where prefix coding is not applied, the new coding scheme uses
artificial vector Gaussian noise as a way of prefix coding.
Moreover, the optimal covariance matrix of the artificial noise
coincides with that of the transmit signal in the previous scheme.
This allows both schemes to be overlayed via S-DPC without
sacrificing the secrecy rate performance for either of them. We
believe that the new understanding of the MIMO Gaussian wiretap
channel problem gained in this work will help to solve some other
multiuser secret communication problems.

\appendices
\section{Proof of Theorem~\ref{thm:MWTC}}\label{app:pf-MWTC}
In this appendix, we prove Theorem~\ref{thm:MWTC}. As mentioned
previously in Remark~\ref{rmk:4}, the secrecy rate on the RHS of
\eqref{eq:SC-LLPS} can be achieved by a coding scheme that combines
Gaussian random binning and prefix coding. We therefore concentrate
on the converse part of the theorem.

Following \cite{LS-ITSubm07}, we will first prove the converse
result for the special case where the channel matrices $\Hv_1$ and
$\Hv_2$ are square and invertible. Next, we will broaden the result
to the general case by approximating arbitrary channel matrices
$\Hv_1$ and $\Hv_2$ by square and invertible ones. For brevity, we
will term the special case as the aligned MIMO Gaussian wiretap
channel and the general case as the general MIMO Gaussian wiretap
channel.

\subsection{Aligned MIMO Gaussian Wiretap Channel}
Consider the special case of the MIMO Gaussian broadcast channel
\eqref{eq:Ch1} where the channel matrices $\Hv_1$ and $\Hv_2$ are
square and invertible. Multiplying both sides of \eqref{eq:Ch1} by
$\Hv_k^{-1}$, the channel model can be equivalently written as
\begin{equation}
\mathbf{Y}_{k}[m] = \mathbf{X}[m]+\mathbf{Z}_k[m], \quad k=1,2
\label{eq:Ch2}
\end{equation}
where $\{\mathbf{Z}_k[m]\}_m$ is an i.i.d. additive vector Gaussian
noise process with zero mean and covariance matrix
\begin{align}
\Nv_k &=\Hv_k^{-1}\Hv_k^{-\T}.
\label{eq:Nk-def}
\end{align}
Denote by $C_s(\Nv_2,\Nv_1,\Sv)$ the secrecy capacity of \eqref{eq:Ch2} (viewed
as a MIMO Gaussian wiretap channel with receiver 2 as legitimate receiver and
receiver 1 as eavesdropper) under the matrix power constraint \eqref{eq:MC}. We
have the following characterization of $C_s(\Nv_2,\Nv_1,\Sv)$.

\begin{lemma}\label{lemma:AMWTC}
The secrecy capacity
\begin{align}
C_s(\Nv_2,\Nv_1,\Sv) &=\max_{0\preceq\Bv\preceq\Sv}\left(
\frac{1}{2}\log\left|\frac{\Sv+\Nv_2}{\Bv+\Nv_2}\right|
-\frac{1}{2}\log\left|\frac{\Sv+\Nv_1}{\Bv+\Nv_1}\right|\right).
\label{eq:SC-LLPS2}
\end{align}
\end{lemma}

\begin{proof}
The achievability of the secrecy rate on the RHS of
\eqref{eq:SC-LLPS2} follows from the achievability of the secrecy
rate on the RHS of \eqref{eq:SC-LLPS} for the general case and the
definition of $\Nv_k$ in \eqref{eq:Nk-def}. To prove the converse
result, we will follow \cite{LS-ITSubm07} and consider a
channel-enhancement argument as follows.

Let us first assume that $\Sv\succ0$. In this case, let $\Bv^\star$ be an
optimal solution to the optimization program on the RHS of \eqref{eq:SC-LLPS2}.
Then, $\Bv^\star$ must satisfy the following Karush-Kuhn-Tucker conditions
\cite{LS-ITSubm07}:
\begin{subequations} \label{eq:KKT}
\begin{align}
(\Bv^{\star}+\Nv_1)^{-1}+\Mv_1&=(\Bv^{\star}+\Nv_2)^{-1}+\Mv_2
\label{eq:KKT1}\\
\Bv^{\star}\Mv_1&=0 \label{eq:KKT2}\\
\text{and}\quad\quad (\Sv-\Bv^{\star})\Mv_2&=0\label{eq:KKT3}
\end{align}
\end{subequations}
where $\Mv_1$ and $\Mv_2$ are positive semidefinite matrices. Let
$\Nt_1$ be a real symmetric matrix such that
\begin{align}
(\Bv^{\star}+\Nt_1)^{-1} &=(\Bv^{\star}+\Nv_1)^{-1}+\Mv_1.
\label{eq:Nt-def}
\end{align}
From Eqns.~(23), (25), (31) and (34) of \cite{LS-ITSubm07}, we have
\begin{align}
0 \prec \Nt_1 \preceq \{\Nv_1,\Nv_2\}, \label{eq:ENH1}
\end{align}
\begin{align}
\left|\frac{\Bv^\star+\Nt_1}{\Nt_1}\right|
=\left|\frac{\Bv^\star+\Nv_1}{\Nv_1}\right| \label{eq:ENH2}
\end{align}
and
\begin{align}
\left|\frac{\Sv+\Nt_1}{\Bv^\star+\Nt_1}\right|
&=\left|\frac{\Sv+\Nv_2}{\Bv^\star+\Nv_2}\right|. \label{eq:ENH3}
\end{align}

Now consider an enhanced MIMO Gaussian broadcast channel:
\begin{align}
\mathbf{Y}_{1}[m] &= \mathbf{X}[m]+\mathbf{Z}_1[m]\notag\\
\text{and} \qquad \mathbf{Y}_{2}[m] &= \mathbf{X}[m]+\tilde{\mathbf{Z}}_1[m]
\label{eq:Ch3}
\end{align}
where $\{\mathbf{Z}_1[m]\}_m$ and $\{\tilde{\mathbf{Z}}_1[m]\}_m$
are i.i.d. additive vector Gaussian noise processes with zero means
and covariance matrices $\Nv_1$ and $\Nt_1$, respectively. Denote by
$C_s(\Nt_1,\Nv_1,\Sv)$ the secrecy capacity of \eqref{eq:Ch3}
(viewed as a MIMO Gaussian wiretap channel with receiver 2 as
legitimate receiver and receiver 1 as eavesdropper) under the matrix
constraint \eqref{eq:MC}. Note from \eqref{eq:ENH1} that
$\Nt_1\preceq\Nv_1$, so the enhanced MIMO Gaussian wiretap channel
\eqref{eq:Ch3} is \emph{degraded}. Hence,
\begin{align}
C_s(\Nt_1,\Nv_1,\Sv) &=
\frac{1}{2}\log\left|\frac{\Sv+\Nt_1}{\Nt_1}\right|
-\frac{1}{2}\log\left|\frac{\Sv+\Nv_1}{\Nv_1}\right|\notag\\
&= \frac{1}{2}\log\left(\left|\frac{\Sv+\Nt_1}{\Sv+\Nv_1}\right|
\left|\frac{\Nv_1}{\Nt_1}\right|\right)\notag\\
&= \frac{1}{2}\log\left(\left|\frac{\Sv+\Nt_1}{\Sv+\Nv_1}\right|
\left|\frac{\Bv^\star+\Nv_1}{\Bv^\star+\Nt_1}\right|\right)\notag\\
&=
\frac{1}{2}\log\left(\left|\frac{\Sv+\Nt_1}{\Bv^\star+\Nt_1}\right|
\left|\frac{\Bv^\star+\Nv_1}{\Sv+\Nv_1}\right|\right)\notag\\
&=
\frac{1}{2}\log\left(\left|\frac{\Sv+\Nv_2}{\Bv^\star+\Nv_2}\right|
\left|\frac{\Bv^\star+\Nv_1}{\Sv+\Nv_1}\right|\right)\notag\\
&= \frac{1}{2}\log\left|\frac{\Sv+\Nv_2}{\Bv^\star+\Nv_2}\right|
-\frac{1}{2}\log\left|\frac{\Sv+\Nv_1}{\Bv^\star+\Nv_1}\right|\label{eq:SC-LS2}
\end{align}
where the first equality follows from \cite[Theorem~1]{LS-ITSubm07};
the third equality follows from \eqref{eq:ENH2}; and the fifth
equality follows from \eqref{eq:ENH3}.

Finally, note from \eqref{eq:ENH1} that $\Nt_1\preceq\Nv_2$, i.e.,
the legitimate receiver in the enhanced wiretap channel
\eqref{eq:Ch3} receives a better signal that the legitimate receiver
in the original wiretap channel \eqref{eq:Ch2}. Therefore,
\begin{align*}
C_s(\Nv_2,\Nv_1,\Sv) &\le C_s(\Nt_1,\Nv_1,\Sv)\\
&= \frac{1}{2}\log\left|\frac{\Sv+\Nv_2}{\Bv^\star+\Nv_2}\right|
-\frac{1}{2}\log\left|\frac{\Sv+\Nv_1}{\Bv^\star+\Nv_1}\right|
\end{align*}
where the last equality follows from \eqref{eq:SC-LS2}. This proved the desired
converse result for $\Sv \succ 0$.

For the case when $\Sv\succeq0$, $|\Sv|=0$, let $$\theta={\sf Rank}(\Sv)<t.$$
Following the same footsteps as in the proof of \cite[Lemma~2]{WSS-IT06}, we
can define an \emph{equivalent} aligned MIMO Gaussian wiretap channel with
$\theta$ transmit and receive antennas and a new covariance matrix power
constraint that is strictly positive definite. Hence, we can convert the case
when $\Sv\succeq0$, $|\Sv|=0$ to the case when $\Sv\succ 0$ with the same
secrecy capacity. This argument can be formally described as follows.

Since $\Sv$ is positive semidefinite, we can write
\begin{align*}
\Sv=\Qv_{\Sv}{\bf \Lambda_{\Sv}}\Qv_{\Sv}^{\T}
\end{align*}
where $\Qv_{\Sv}$ is an orthogonal matrix and
\begin{align*}
{\bf \Lambda_{\Sv}}
=\diag(\underbrace{0,\dots,0}_{t-\theta},s_1,\dots,s_{\theta})
\end{align*}
is diagonal with $s_j>0$, $j=1,\dots,\theta$. For $k=1,2$, write
\begin{align*}
\Qv_{\Sv}^{\T}\Nv_k \Qv_{\Sv}= \left(
\begin{matrix} \Cv_k & \Dv_k \\ \Dv_k^{\T} & \Ev_k\end{matrix}\right)
\end{align*}
where $\Cv_k$, $\Dv_k$ and $\Ev_k$ are (sub)matrices of size
$(t-\theta)\times(t-\theta)$, $(t-\theta)\times\theta$ and
$\theta\times\theta$, respectively. Let
\begin{align*}
\Av_k := \left(
\begin{matrix} \Iv_{t-\theta} & 0_{(t-\theta)\times \theta}
\\ -\Dv_k^{\T}\Cv_k^{-1} & \Iv_{\theta}
\end{matrix}\right), \quad k=1,2.
\end{align*}

We now define an intermediate and equivalent channel by multiplying both sides
of \eqref{eq:Ch2} by an \emph{invertible} matrix $\Av_k\Qv_{\Sv}^{\T}$:
\begin{align}
\Yv_k'[m] &= \Xv'[m]+ \Zv_k'[m], \quad k=1,2 \label{eq:Ch4}
\end{align}
where
\begin{align*}
\Yv_k'[m] &= \Av_k\Qv_{\Sv}^{\T}\Yv_k[m]\\
\Xv'[m] &= \Av_k\Qv_{\Sv}^{\T}\Xv[m]\\
\mbox{and} \quad\quad \Zv_k'[m] &= \Av_k\Qv_{\Sv}^{\T}\Zv_k[m].
\end{align*}
Then, the covariance matrix $\Nv_k'$ of the additive Gaussian noise
vector $\Zv_k'[m]$ is given by
\begin{align}
\Nv_k' &=\left(
\begin{matrix}
\Cv_k & 0 \\ 0 &
\Ev_k-\Dv_k^{\T}\Cv_k^{-1}\Dv_k
\end{matrix}\right).\label{eq:Nk'-def}
\end{align}
and the matrix power constraint \eqref{eq:MC} becomes
\begin{align}
\frac{1}{n}\sum_{m=1}^{n}\mathbf{X}'[m]{\mathbf{X}'}^{\T}[m] \preceq
\mathbf{S}' \label{eq:MC2}
\end{align}
where
\begin{align}
\Sv'&=\Av_k\Qv_{\Sv}^{\T}\Sv\Qv_{\Sv}\Av_k^{\T}\notag\\
&=\Av_k {\bf \Lambda_{\Sv}} \Av_k^{\T}\notag\\
&={\bf \Lambda_{\Sv}}.\label{eq:S'-def}
\end{align}

Note from \eqref{eq:S'-def} that $\Sv'$ is diagonal with first
$t-\theta$ diagonal elements equal to zero. Thus, the matrix
constraint \eqref{eq:MC2} requires that the first $t-\theta$
elements of $\Xv'[m]$ be zero. Moreover, from \eqref{eq:Nk'-def},
the first $t-\theta$ and the rest of $\theta$ elements of
$\Zv'_k[m]$ are uncorrelated and hence must be independent as
$\Zv'_k[m]$ is Gaussian. Therefore, only the latter $\theta$
antennas transmit/receive information regarding message $W$. This
allows us to define another \emph{equivalent} aligned MIMO Gaussian
broadcast channel with $\theta$ antennas at the transmitter and each
of the receivers:
\begin{align}
\Ybv_k[m] &= \Xbv[m]+\Zbv_k[m], \quad k=1,2 \label{eq:Ch5}
\end{align}
where
\begin{align*}
\Ybv_k[m] &= \Abv\Yv'_k[m]\\
\Xbv[m] &= \Abv\Xv'[m]\\
\Zbv_k[m] &= \Abv\Zv'_k[m]
\end{align*}
and $\Abv=\left[0_{\theta\times (t-\theta)}\;\Iv_{\theta}\right]$. Now, the
matrix power constraint \eqref{eq:MC2} becomes
\begin{align}
\frac{1}{n}\sum_{m=1}^{n}\Xbv[m]\Xbv^{\T}[m] \preceq \Sbv
\label{eq:MC3}
\end{align}
where
\begin{align}
\Sbv &=\Abv\Sv'\Abv^{\T}\notag\\
&=\diag(s_1,\dots,s_{\theta}).\label{eq:Sb-def}
\end{align}
Note that the matrix power constraint $\Sbv$ is \emph{strictly} positive
definite, so we can apply the previous result to the new wiretap channel
\eqref{eq:Ch5}. This completes the proof of the lemma.
\end{proof}

\subsection{General MIMO Gaussian Wiretap Channel}
For the general case, we may assume that the channel matrices $\Hv_1$ and
$\Hv_2$ are square but not necessarily invertible. If that is not the case, we
can use singular value decomposition (SVD) to show that there is an equivalent
channel which does have $t \times t$ square channel matrices. That is, we can
find a new channel with square channel matrices which are derived from the
original ones via matrix multiplications. The new channel is equivalent to the
original one in preserving the secrecy capacity under the same power
constraint.

Consider using SVD to write the channel matrices as follows:
$$\mathbf{H}_k=\Uv_k\boldsymbol{\Lambda}_k\Vv_k^\T, \quad
k=1,2$$ where $\Uv_k$ and $\Vv_k$ are $t \times t$ orthogonal matrices, and
$\boldsymbol{\Lambda}_k$ is diagonal. We now define a new MIMO Gaussian
broadcast channel which has invertible channel matrices:
\begin{align}
\Yv_k[m] &= \overline{\Hv}_k\Xv[m]+\Zv_k[m], \quad k=1,2
\label{eq:Ch6}
\end{align}
where
$$\overline{\mathbf{H}}_k=\Uv_k(\boldsymbol{\Lambda}_k+\alpha\mathbf{I}_t)\Vv_k^t$$
for some $\alpha > 0$, and $\{\mathbf{Z}_k[m]\}_m$ is an i.i.d. additive vector
Gaussian noise process with zero mean and identity covariance matrix. Note that
the channel matrices $\overline{\Hv}_k$, $k=1,2$, are invertible. By
Lemma~\ref{lemma:AMWTC}, the secrecy capacity
$C_s(\overline{\mathbf{H}}_2,\overline{\mathbf{H}}_1,\Sv)$ of \eqref{eq:Ch2}
(viewed as a MIMO Gaussian wiretap channel with receiver 2 as legitimate
receiver and receiver 1 as eavesdropper) under the matrix power constraint
\eqref{eq:MC} is given by
\begin{equation*}
C_s(\overline{\mathbf{H}}_2,\overline{\mathbf{H}}_1,\Sv) =
\max_{0\preceq\Bv\preceq\mathbf{S}}
\left(\frac{1}{2}\log\left|\frac{\Iv_t+\overline{\Hv}_2\Sv\overline{\Hv}_2^{\T}}
{\Iv_t+\overline{\Hv}_2\Bv\overline{\Hv}_2^{\T}}\right|
-\frac{1}{2}\log\left|\frac{\Iv_t+\overline{\Hv}_1\Sv\overline{\Hv}_1^{\T}}
{\Iv_t+\overline{\Hv}_1\Bv\overline{\Hv}_1^{\T}}\right|\right).
\end{equation*}

Finally, let $\alpha\downarrow0$. We have $\Hbv_k\rightarrow\Hv_k$,
$k=1,2$ and hence
\begin{align*}
C_s(\overline{\mathbf{H}}_2,\overline{\mathbf{H}}_1,\Sv) \rightarrow
\max_{0\preceq\Bv\preceq\Sv}\left(
\frac{1}{2}\log\left|\frac{\Iv_{r_2}+\Hv_2\Sv\Hv_2^{\T}}{\Iv_{r_2}+\Hv_2\Bv\Hv_2^{\T}}\right|
-\frac{1}{2}\log\left|\frac{\Iv_{r_1}+\Hv_1\Sv\Hv_1^{\T}}{\Iv_{r_1}+\Hv_1\Bv\Hv_1^{\T}}\right|\right).
\end{align*}
Moreover, by Eqns.~(45) and (46) of \cite{LS-ITSubm07},
\begin{align}
C_s(\mathbf{H}_2,\mathbf{H}_1,\Sv) &\le
C_s(\overline{\mathbf{H}}_2,\overline{\mathbf{H}}_1,\Sv)+\mathcal{O}(\alpha)
\label{eq:SC-LLPS3}
\end{align}
where $\mathcal{O}(\alpha)\rightarrow 0$ in the limit as
$\alpha\downarrow0$. Thus, we have the desired converse result
\begin{align*}
C_s(\mathbf{H}_2,\mathbf{H}_1,\Sv) &\le
\max_{0\preceq\Bv\preceq\Sv}\left(
\frac{1}{2}\log\left|\frac{\Iv_{r_2}+\Hv_2\Sv\Hv_2^{\T}}{\Iv_{r_2}+\Hv_2\Bv\Hv_2^{\T}}\right|
-\frac{1}{2}\log\left|\frac{\Iv_{r_1}+\Hv_1\Sv\Hv_1^{\T}}{\Iv_{r_1}+\Hv_1\Bv\Hv_1^{\T}}\right|\right)
\end{align*}
by letting $\alpha\downarrow0$ on the RHS of \eqref{eq:SC-LLPS3}.
This completes the proof of the theorem.

\section{Proof of Theorem~\ref{thm:MWTC2}} \label{app:pf-MWTC2}
In this appendix, we prove Theorem~\ref{thm:MWTC2}. Without loss of
generality, we may assume that the matrix power constraint $\Sv$ is
strictly positive definite and the channel matrices $\Hv_1$ and
$\Hv_2$ are square but not necessarily invertible. We start with the
following simple lemma.

\begin{lemma} \label{lem:cal}
For any $t \times t$ matrices $\Bv$ and $\Hv$ such that
$\Bv\succeq0$, we have
\begin{align}
\left|\Iv_{t}+\Hv\Bv\Hv^{\T}\right|&=\left|\Iv_{t}+\Hv^{\T}\Hv\Bv\right|.
\label{eq:cal1}
\end{align}
In particular, if $\Bv=\Iv_t$, we have
\begin{align}
\left|\Iv_{t}+\Hv\Hv^{\T}\right|&=\left|\Iv_{t}+\Hv^{\T}\Hv\right|.
\label{eq:cal2}
\end{align}
\end{lemma}

\begin{proof} Note that if $\Hv$ is invertible, the equalities in
\eqref{eq:cal1} and \eqref{eq:cal2} are trivial. Otherwise, consider
using SVD to rewrite $\Hv$ as
\begin{align*}
\mathbf{H}=\Uv\boldsymbol{\Lambda}\Vv^\T
\end{align*}
where $\Uv$ and $\Vv$ are $t \times t$ orthogonal matrices, and
\begin{align*}
{\bf \Lambda}
=\diag(\underbrace{0,\dots,0}_{t-b},\lambda_1,\dots,\lambda_{b})
\end{align*}
is diagonal with $\lambda_j>0$, $j=1,\dots,b$. Write
\begin{align*}
\Vv^\T\Bv \Vv= \left(
\begin{matrix} \Cv_{\Bv} & \Dv_{\Bv} \\ \Dv_{\Bv}^{\T} & \Ev_{\Bv}\end{matrix}\right)
\end{align*}
where $\Cv_{\Bv}$, $\Dv_{\Bv}$ and $\Ev_{\Bv}$ are (sub)matrices of
size $(t-b)\times(t-b)$, $(t-b)\times b$ and $b \times b$,
respectively. Then,
\begin{align}
\left|\Iv_t+\Hv\Bv\Hv^{\T}\right|
&=\left|\Iv_t+\Uv\boldsymbol{\Lambda}\Vv^\T\Bv\Vv\boldsymbol{\Lambda}\Uv^{\T}\right|\notag\\
&=\left|\Iv_t+ \boldsymbol{\Lambda}\Vv^\T\Bv\Vv\boldsymbol{\Lambda}
\right| \notag\\
&=\left|\Iv_{b}+ \overline{\boldsymbol{\Lambda}}\Ev_{\Bv}
\overline{\boldsymbol{\Lambda}} \right| \label{eq:cal3}
\end{align}
where
$\overline{\boldsymbol{\Lambda}}=\diag(\lambda_1,\dots,\lambda_{b})$.
On the other hand,
\begin{align}
\left|\Iv_t+\Hv^{\T}\Hv\Bv\right|
&=\left|\Iv_t+\Vv\boldsymbol{\Lambda}^2\Vv^\T\Bv\right|\notag\\
&=\left|\Iv_t+\boldsymbol{\Lambda}^2\Vv^\T\Bv\Vv\right|\notag\\
&=\left|\Iv_{b}+\overline{\boldsymbol{\Lambda}}^2\Ev_{\Bv}\right|\notag\\
&=\left|\Iv_{b}+\overline{\boldsymbol{\Lambda}}\Ev_{\Bv}
\overline{\boldsymbol{\Lambda}}\right| \label{eq:cal4}
\end{align}
where the last equality follows from the fact that
$\overline{\boldsymbol{\Lambda}}$ is invertible. Putting together
\eqref{eq:cal3} and \eqref{eq:cal4} proves the equality in
\eqref{eq:cal1}. This completes the proof of the lemma.
\end{proof}

We are now ready to prove Theorem~\ref{thm:MWTC2}, following the
approach of \cite{BLVS-EURASIP08}. Let
\begin{align}
\Ov_k : =\Hv_k^{\T}\Hv_k  \quad k=1,2, \label{eq:def-Ok}
\end{align}
and let $\boldsymbol{\Phi}$ denote the generalized eigenvalue matrix
of the pencil
$$\left(\Iv_t+\Sv^{\frac{1}{2}}\Ov_1\Sv^{\frac{1}{2}},\;
\Iv_t+\Sv^{\frac{1}{2}}\Ov_2\Sv^{\frac{1}{2}}\right)$$
such that
\begin{align*}
\boldsymbol{\Phi}=\left(\begin{matrix} \overline{\boldsymbol{\Phi}}_1 & 0 \\
0 & \overline{\boldsymbol{\Phi}}_2\end{matrix}\right)
\end{align*}
where $\overline{\boldsymbol{\Phi}}_1={\rm
Diag}\{\phi_{1},\dots,\phi_{\rho}\}$ and
$\overline{\boldsymbol{\Phi}}_2 ={\rm
Diag}\{\phi_{\rho+1},\dots,\phi_{t}\}$. Let $\Gv$ be the
corresponding generalized eigenvector matrix such that
\begin{align}
\Gv^{\T}\left(\Iv_t+\Sv^{\frac{1}{2}}\Ov_1\Sv^{\frac{1}{2}}\right)\Gv &=\boldsymbol{\Phi} \notag\\
\text{and}\qquad
\Gv^{\T}\left(\Iv_t+\Sv^{\frac{1}{2}}\Ov_2\Sv^{\frac{1}{2}}\right)\Gv&=\Iv_t.
\label{eq:def-G}
\end{align}
Now define
\begin{align}
\Ot:= \Sv^{-\frac{1}{2}}\left[ \Gv^{-\T}
\left(\begin{matrix} \overline{{\boldsymbol{\Phi}}}_1 & 0 \\
0& \Iv_{t-\rho}\end{matrix}\right)\Gv^{-1}-\Iv_t \right]\Sv^{-\frac{1}{2}}.
\label{eq:def-O}
\end{align}
Since the generalized eigenvalues are ordered as
$$\phi_{1}\ge\dots \ge \phi_{\rho}> 1 \ge \phi_{\rho+1}\ge\dots\ge \phi_{t}>0,$$
we have
\begin{align*}
\left(\begin{matrix} \overline{{\boldsymbol{\Phi}}}_1 & 0 \\
0& \Iv_{t-\rho}\end{matrix}\right)\succeq \boldsymbol{\Phi}\\
\mbox{and} \quad\quad
\left(\begin{matrix} \overline{{\boldsymbol{\Phi}}}_1 & 0 \\
0& \Iv_{t-\rho}\end{matrix}\right)\succeq \Iv_t.
\end{align*}
Hence by \eqref{eq:def-G} and \eqref{eq:def-O},
\begin{align}
\Ot \succeq \{\Ov_1,\Ov_2\}. \label{eq:O-order}
\end{align}
It follows that
\begin{align}
C_s(\Hv_1,\Hv_2,\Sv) &= \max_{0\preceq\Bv\preceq\Sv}\left(
\frac{1}{2}\log\left|\Iv_{t}+\Hv_1\Bv\Hv_1^\T\right|-
\frac{1}{2}\log\left|\Iv_{t}+\Hv_2\Bv\Hv_2^\T\right|\right)\notag\\
&=\max_{0\preceq\Bv\preceq\Sv}\left(
\frac{1}{2}\log\left|\Iv_{t}+\Bv^\frac{1}{2}\Hv_1^\T\Hv_1\Bv^\frac{1}{2}\right|-
\frac{1}{2}\log\left|\Iv_{t}+\Bv^\frac{1}{2}\Hv_2^\T\Hv_2\Bv^\frac{1}{2}\right|\right)\label{1}\\
&=\max_{0\preceq\Bv\preceq\Sv}\left(
\frac{1}{2}\log\left|\Iv_{t}+\Bv^\frac{1}{2}\Ov_1\Bv^\frac{1}{2}\right|-
\frac{1}{2}\log\left|\Iv_{t}+ \Bv^\frac{1}{2}\Ov_2\Bv^\frac{1}{2}\right|\right)\label{2}\\
&\le \max_{0\preceq\Bv\preceq\Sv}\left(
\frac{1}{2}\log\left|\Iv_t+\Bv^\frac{1}{2}\Ot\Bv^\frac{1}{2}\right|-
\frac{1}{2}\log\left|\Iv_t+\Bv^\frac{1}{2}\Ov_2\Bv^\frac{1}{2}\right|\right)\label{3}\\
&= \max_{0\preceq\Bv\preceq\Sv}\left(
\frac{1}{2}\log\left|\Iv_t+\Ot^\frac{1}{2}\Bv\Ot^\frac{1}{2}\right|-
\frac{1}{2}\log\left|\Iv_t+\Ov_2^\frac{1}{2}\Bv\Ov_2^\frac{1}{2}\right|\right)\label{4}\\
&=
\frac{1}{2}\log\left|\Iv_t+\Ot^\frac{1}{2}\Sv\Ot^\frac{1}{2}\right|-
\frac{1}{2}\log\left|\Iv_t+\Ov_2^\frac{1}{2}\Sv\Ov_2^\frac{1}{2}\right|\label{5}\\
&= \frac{1}{2}\log\left|\Iv_t+\Sv^{\frac{1}{2}} \Ot\Sv^{\frac{1}{2}}
\right|-\frac{1}{2}\log\left|\Iv_t+ \Sv^{\frac{1}{2}}\Ov_2
\Sv^{\frac{1}{2}}\right|\label{6}\\
&=\frac{1}{2}\log\left|\overline{{\boldsymbol{\Phi}}}_1\right|\label{7}\\
&=\frac{1}{2} \sum_{j=1}^{\rho} \log \phi_{j} \label{eq:cmpb-U}
\end{align}
where \eqref{1}, \eqref{4} and \eqref{6} follow from
\eqref{eq:cal2}; \eqref{2} follows from the definition of $\Ov_1$ in
\eqref{eq:def-Ok}; \eqref{3} follows from the fact that
$\Ov_1\preceq\Ot$ (see \eqref{eq:O-order}); \eqref{5} follows from
the fact that $\Ov_2\preceq\Ot$ (see \eqref{eq:O-order}); and
\eqref{7} follows \eqref{eq:def-G} and the definition of $\Ot$ in
\eqref{eq:def-O}.

To prove the reverse inequality, let $\Gv=[\Gv_1\, \Gv_2]$ where $\Gv_1$ and
$\Gv_2$ are (sub)matrices of size $t\times \rho$ and $t\times \rho$,
respectively, and let
\begin{align}
\Bv^{\star} := \Sv^{\frac{1}{2}}\Gv
\left(\begin{matrix} \left(\Gv_1^{\T}\Gv_1\right)^{-1} & 0 \\
0& 0 \end{matrix}\right)\Gv^{\T}\Sv^{\frac{1}{2}}.
\end{align}
Then, $\Bv^\star$ is positive semidefinite. Moreover, we may verify
that $\Bv^{\star}\preceq \Sv$ as follows. Note that $\Gv$ is
invertible, so it is enough to show that
\begin{align*}
\left(\begin{matrix} (\Gv_1^{\T}\Gv_1)^{-1} & 0 \\
0& 0 \end{matrix}\right) \preceq \left(\Gv^{\T}\Gv\right)^{-1}.
\end{align*}
Note that
\begin{align*}
\Gv^{\T}\Gv &= \left(\begin{matrix} \Gv_1^{\T}\Gv_1 & \Gv_1^{\T}\Gv_2 \\
\Gv_2^{\T}\Gv_1 & \Gv_2^{\T}\Gv_2 \end{matrix}\right).
\end{align*}
Using block inversion, we may obtain
\begin{align*}
\left(\Gv^{\T}\Gv\right)^{-1} &= \left(\begin{matrix} (\Gv_1^{\T}\Gv_1)^{-1} +
(\Gv_1^{\T}\Gv_1)^{-1}\Gv_1^{\T}\Gv_2  \Ev_{\Gv}^{-1}  \Gv_2^{\T}\Gv_1
(\Gv_1^{\T}\Gv_1)^{-1} & (\Gv_1^{\T}\Gv_1)^{-1}\Gv_1^{\T}\Gv_2\Ev_{\Gv}^{-1} \\
\Ev_{\Gv}^{-1}\Gv_2^{\T}\Gv_1 (\Gv_1^{\T}\Gv_1)^{-1} & \Ev_{\Gv}^{-1}
\end{matrix}\right)
\end{align*}
where
\begin{align*}
\Ev_{\Gv}=\Gv_2^{\T}\Gv_2 - \Gv_2^{\T}\Gv_1
(\Gv_1^{\T}\Gv_1)^{-1}\Gv_1^{\T}\Gv_2.
\end{align*}
Since $\Gv^{\T}\Gv$ is positive definite, we have
$$\Ev_{\Gv} \succ 0$$ and hence
\begin{align*}
\left(\Gv^{\T}\Gv\right)^{-1} - \left(\begin{matrix} (\Gv_1^{\T}\Gv_1)^{-1} & 0 \\
0& 0 \end{matrix}\right)  &= \left(\begin{matrix}
(\Gv_1^{\T}\Gv_1)^{-1}\Gv_1^{\T}\Gv_2  \Ev_{\Gv}^{-1}  \Gv_2^{\T}\Gv_1
(\Gv_1^{\T}\Gv_1)^{-1} & (\Gv_1^{\T}\Gv_1)^{-1}\Gv_1^{\T}\Gv_2\Ev_{\Gv}^{-1} \\
\Ev_{\Gv}^{-1}\Gv_2^{\T}\Gv_1 (\Gv_1^{\T}\Gv_1)^{-1} & \Ev_{\Gv}^{-1} \end{matrix}\right)\\
&=\left(\begin{matrix} \Iv_{\rho} & (\Gv_1^{\T}\Gv_1)^{-1}\Gv_1^{\T}\Gv_2 \\
0 & \Iv_{t-\rho} \end{matrix}\right)
\left(\begin{matrix} 0 & 0 \\
0 & \Ev_{\Gv}^{-1} \end{matrix}\right)
\left(\begin{matrix} \Iv_{\rho} & 0 \\
\Gv_2^{\T}\Gv_1 (\Gv_1^{\T}\Gv_1)^{-1} & \Iv_{t-\rho} \end{matrix}\right) \\
&\succeq 0.
\end{align*}

By \eqref{2},
\begin{align}
C_s(\Hv_1,\Hv_2,\Sv) &=\max_{0\preceq\Bv\preceq\Sv}\left(
\frac{1}{2}\log\left|\Iv_{t}+\Bv^\frac{1}{2}\Ov_1\Bv^\frac{1}{2}\right|-
\frac{1}{2}\log\left|\Iv_{t}+ \Bv^\frac{1}{2}\Ov_2\Bv^\frac{1}{2}\right|\right)\notag\\
&\ge
\frac{1}{2}\log\left|\Iv_{t}+{\Bv^\star}^\frac{1}{2}\Ov_1{\Bv^\star}^\frac{1}{2}\right|-
\frac{1}{2}\log\left|\Iv_{t}+{\Bv^\star}^\frac{1}{2}\Ov_2{\Bv^\star}^\frac{1}{2}\right|\notag\\
&= \frac{1}{2}\log\left|\Iv_{t}+{\Bv^\star}\Ov_1\right|-
\frac{1}{2}\log\left|\Iv_{t}+{\Bv^\star}\Ov_2\right|\label{eq:cmpb-L1}
\end{align}
where the last equality follows from \eqref{eq:cal1}. From
\eqref{eq:def-G}, we have
\begin{align}
\Ov_1 &= \Sv^{-\frac{1}{2}}\left(\Gv^{-\T}
\boldsymbol{\Phi}\Gv^{-1}-\Iv_t
\right)\Sv^{-\frac{1}{2}}\notag\\
\text{and}\qquad \Ov_2 &=
\Sv^{-\frac{1}{2}}\left(\Gv^{-\T}\Gv^{-1}-\Iv_t
\right)\Sv^{-\frac{1}{2}}.
\end{align}
Hence,
\begin{align*}
\Bv^{\star} \Ov_1 &= \Sv^\frac{1}{2}\Gv
\left(\begin{matrix} \left(\Gv_1^{\T}\Gv_1\right)^{-1} & 0 \\
0& 0 \end{matrix}\right)\Gv^{\T} \left(\Gv^{-\T}
\boldsymbol{\Phi}\Gv^{-1}-\Iv_t\right)\Sv^{-\frac{1}{2}}\\
&= \Sv^\frac{1}{2}\Gv\left[
\left(\begin{matrix} \left(\Gv_1^{\T}\Gv_1\right)^{-1} & 0 \\
0& 0 \end{matrix}\right)\boldsymbol{\Phi}-\left(\begin{matrix} \left(\Gv_1^{\T}\Gv_1\right)^{-1} & 0 \\
0& 0 \end{matrix}\right)\Gv^{\T}\Gv\right] \Gv^{-1}\Sv^{-\frac{1}{2}}\\
&= \Sv^\frac{1}{2}\Gv\left[
\left(\begin{matrix} \left(\Gv_1^{\T}\Gv_1\right)^{-1} & 0 \\
0& 0 \end{matrix}\right)\left(\begin{matrix} \overline{\boldsymbol{\Phi}}_1 & 0 \\
0 & \overline{\boldsymbol{\Phi}}_2\end{matrix}\right)-\left(\begin{matrix} \left(\Gv_1^{\T}\Gv_1\right)^{-1} & 0 \\
0& 0 \end{matrix}\right)
\left(\begin{matrix} \Gv_1^{\T}\Gv_1 & \Gv_1^{\T}\Gv_2 \\
\Gv_2^{\T}\Gv_1 & \Gv_2^{\T}\Gv_2 \end{matrix}\right)
\right]\Gv^{-1}\Sv^{-\frac{1}{2}}\\
&= \Sv^\frac{1}{2}\Gv\left(\begin{matrix}
\left(\Gv_1^{\T}\Gv_1\right)^{-1}\overline{\boldsymbol{\Phi}}_1-\Iv_\rho
&
-\left(\Gv_1^{\T}\Gv_1\right)^{-1}\Gv_1^{\T}\Gv_2 \\
0& 0 \end{matrix}\right)\Gv^{-1}\Sv^{-\frac{1}{2}}
\end{align*}
giving
\begin{align}
\left|\Iv_{t}+{\Bv^\star}\Ov_1\right|=\left|\Gv_1^{\T}\Gv_1\right|^{-1}
\left|\overline{\boldsymbol{\Phi}}_1\right|. \label{eq:cmpb-L2}
\end{align}
Similarly, we may obtain
\begin{align*}
\Bv^{\star} \Ov_2 &= \Sv^\frac{1}{2}\Gv\left(\begin{matrix}
\left(\Gv_1^{\T}\Gv_1\right)^{-1}-\Iv_\rho &
-\left(\Gv_1^{\T}\Gv_1\right)^{-1}\Gv_1^{\T}\Gv_2 \\
0& 0 \end{matrix}\right)\Gv^{-1}\Sv^{-\frac{1}{2}}
\end{align*}
and
\begin{align}
\left|\Iv_{t}+\Bv^{\star} \Ov_2\right| &=
\left|\Gv_1^{\T}\Gv_1\right|^{-1}. \label{eq:cmpb-L3}
\end{align}
Substituting \eqref{eq:cmpb-L2} and \eqref{eq:cmpb-L3} into
\eqref{eq:cmpb-L1}, we may obtain
\begin{align}
C_s(\Hv_1,\Hv_2,\Sv) & \ge  \frac{1}{2}\log\left|\overline{\boldsymbol{\Phi}}_1\right|\notag\\
&=\frac{1}{2} \sum_{j=1}^{\rho} \log \phi_{j}. \label{eq:cmpb-L4}
\end{align}
Putting together \eqref{eq:cmpb-U} and \eqref{eq:cmpb-L4}
establishes the desired equality
\begin{align*}
C_s(\Hv_1,\Hv_2,\Sv) &=\frac{1}{2} \sum_{j=1}^{\rho} \log \phi_{j}.
\end{align*}
This completes the proof of the theorem.

\bibliographystyle{IEEEtran}
\bibliography{secrecy}

\end{document}